\documentclass[twocolumn,aps,prb,superscriptaddress, amsmath,amssymb,longbibliography,prb]{revtex4-2}
\newcommand{\figurewidth}{.9\columnwidth}

\usepackage{graphicx}
\usepackage{siunitx}	
\usepackage{amssymb}
\usepackage{datetime}
\usepackage{booktabs}
\usepackage{hyperref}
\usepackage{xcolor}
\hypersetup{
    colorlinks,
    linkcolor={red!50!black},
    citecolor={blue!50!black},
    urlcolor={blue!80!black}
}

\usepackage{todonotes}
\newcommand{\HS}{\texorpdfstring{H$_3$S}{H3S}}
\newcommand{\YHix}{\texorpdfstring{YH$_9$}{YH9}}
\newcommand{\HiiS}{\texorpdfstring{H$_2$S}{H2S}}
\newcommand{\Hcii}{\ensuremath{H_{\text c2}}}
\newcommand{\Bcii}{\ensuremath{B_{\text c2}}}

\newcommand{\GPa}{\giga\pascal}
\newcommand{\um}{\micro\meter}
\newcommand{\dd}{\text{d}}	
\newcommand{\Tc}{\ensuremath{T_{\text c}}}

\newcommand{\Imtm}{$Im\bar{3}m$}
\newcommand{\ThetaD}{\ensuremath{\Theta_{\text D}}}

\newcommand{\kB}{\ensuremath{k_{\text B}}}

\newcommand{\vF}{\ensuremath{v_{\text F}}}

\DeclareMathOperator{\sgn}{sgn}
\DeclareSIUnit\rydberg{Ry}

\graphicspath{{Figures/}} 

\begin{document}

%
%

\preprint{ver 0.10 \today\ \currenttime}

\title{Clean-limit superconductivity in \Imtm\ \HS\ synthesized from sulfur and hydrogen donor ammonia borane}

\author{Israel Osmond}
\affiliation{H.H. Wills Physics Laboratory, University of Bristol, Tyndall Avenue, Bristol, BS8 1TL, UK}

\author{Owen Moulding}
\affiliation{H.H. Wills Physics Laboratory, University of Bristol, Tyndall Avenue, Bristol, BS8 1TL, UK}

\author{Sam Cross}
\affiliation{H.H. Wills Physics Laboratory, University of Bristol, Tyndall Avenue, Bristol, BS8 1TL, UK}

\author{Takaki Muramatsu}
\affiliation{H.H. Wills Physics Laboratory, University of Bristol, Tyndall Avenue, Bristol, BS8 1TL, UK}

\author{Annabelle Brooks}
\affiliation{H.H. Wills Physics Laboratory, University of Bristol, Tyndall Avenue, Bristol, BS8 1TL, UK}

\author{Oliver Lord}
\affiliation{School of Earth Sciences, University of Bristol, Wills Memorial Building, Queen’s Road,
Bristol BS8 1RJ, United Kingdom}

\author{Timofey Fedotenko}
\affiliation{ Photon Science, DESY, Notkestrasse 85, 22607 Hamburg, Germany}

\author{Jonathan Buhot}
\affiliation{H.H. Wills Physics Laboratory, University of Bristol, Tyndall Avenue, Bristol, BS8 1TL, UK}

\author{Sven Friedemann}
\affiliation{H.H. Wills Physics Laboratory, University of Bristol, Tyndall Avenue, Bristol, BS8 1TL, UK}
\email{Sven.Friedemann@bristol.ac.uk}
	
\date{\today}

\begin{abstract}
We present detailed studies of the superconductivity in high-pressure \HS. X-ray diffraction measurements show that cubic \Imtm\ \HS\ was synthesized from elemental sulfur and hydrogen donor ammonia borane (NH$_3$BH$_3$). Our electrical transport measurements confirm superconductivity with a transition temperature $\Tc = \SI{197}{\kelvin}$ at \SI{153}{\GPa}. From the analysis of both the normal state resistivity and the slope of the critical field, we conclude that the superconductivity is described by clean-limit behaviour. A significant broadening of the resistive transition in finite magnetic field is found, as expected for superconductors. We identify a linear temperature-over-field scaling of the resistance at the superconducting transition which is not described by existing theories.
\end{abstract}



\maketitle

The discovery of superconductivity in \HS\ at a critical temperature of $\Tc\sim\SI{200}{\kelvin}$  revolutionized the search for high-temperature superconductivity \cite{Drozdov2015}. Since then, high-temperature superconductivity has been observed in a number of binary hydrides at high pressures including LaH$_{10}$, CeH$_9$, and YH$_{9}$ and in carbonaceous sulfur hydride \cite{Drozdov2015,Chen2021b,Snider2020,Somayazulu2019,Kong2021,Drozdov2019}. Extreme pressures of typically more than \SI{100}{\GPa} are required for the synthesis and study of these hydride compounds. The high hydrogen stoichiometries yield the high density of electronic states, high-frequency phonon modes, and strong electron-phonon coupling necessary for a high-\Tc\ in phonon mediated superconductivity \cite{Drozdov2019, Kong2021,Errea2016,Errea2020,Pickard2020}. Both synthesis and experimental studies of high-temperature hydride superconductors remain very challenging due to the high pressures needed. 

Synthesis of hydride superconductors is typically done by laser heating precursors \textit{in situ} at high pressures. Only in the initial work by Drozdov \textit{et al.} was superconducting \HS\ synthesized from the dissociation of molecular H$_2$S at high pressures \cite{Drozdov2015,Goncharov2016,Li2016,Huang2019}. All recent studies  synthesized  \HS{} from elemental sulfur and either molecular hydrogen or a hydrogen donor material as precursors using laser heating \cite{Mozaffari2019,Pace2020,Guigue2017,Nakao2019,Snider2020,Snider2021,Minkov2021,Laniel2020}. 
Whilst molecular hydrogen affords the cleanest synthesis route, loading hydrogen into a diamond anvil cell (DAC) is technically much more demanding. Thus, it is not easily adapted for wide-spread and detailed studies of superconductivity in hydrides by the global physics community. Hence, it is important to establish synthesis routes using hydrogen donor materials like ammonia borane. For some hydrides this can be further simplified with  evaporated samples of starting elements such as yttrium or lanthanum \cite{Snider2021,Buhot2020}.

Detailed understanding of the superconductivity in hydride compounds requires structural information, e.g. from x-ray diffraction (XRD) to complement information about the superconducting properties -- ideally on the same sample. 
Transition temperatures up to \SI{203}{\kelvin}  have been linked to the  cubic \Imtm\ phase of \HS\ at a pressure of \SI{155}{\GPa} \cite{Drozdov2015,Einaga2016}. At pressures below $\approx\SI{140}{\GPa}$ a rhombohedral distortion leads to a lower symmetry $R3m$ phase with a reduced \Tc\ \cite{Goncharov2017}. Other phases have been reported but have not been probed for superconductivity \cite{Goncharov2016,Li2016,Guigue2017,Laniel2020,Pace2020}.

Superconductivity in \HS\ has been confirmed with multiple probes despite the limitations and challenges of measurements in DACs. The most common evidence stems from the observation of zero resistance
for various samples by the group of Eremets and the suppression of the resistive transition in magnetic field \cite{Drozdov2015,Einaga2016,Mozaffari2019}. In addition, a suppression of \Tc\ has been observed for samples with deuterium substitution \cite{Drozdov2015,Einaga2016} roughly in agreement with the expected isotope effect predicted by computational studies \cite{Errea2016,VillaCortes2022}. A diamagnetic signal has been observed in DC magnetisation measurements \cite{Drozdov2015,Minkov2021}, nuclear resonant scattering \cite{Troyan2016}, and in AC susceptibility by \cite{Huang2019}.
Recently, Minkov \textit{et al.}  have presented detailed magnetisation studies of \HS\ synthesized from sulfur and ammonia borane from which they extract the lower critical field and London penetration depth\cite{Minkov2021}. Here, we present electrical resistance measurements demonstrating superconductivity in \HS\ synthesized with this novel route using ammonia borane and sulfur. 

\begin{figure*}%
\includegraphics[width=\textwidth]{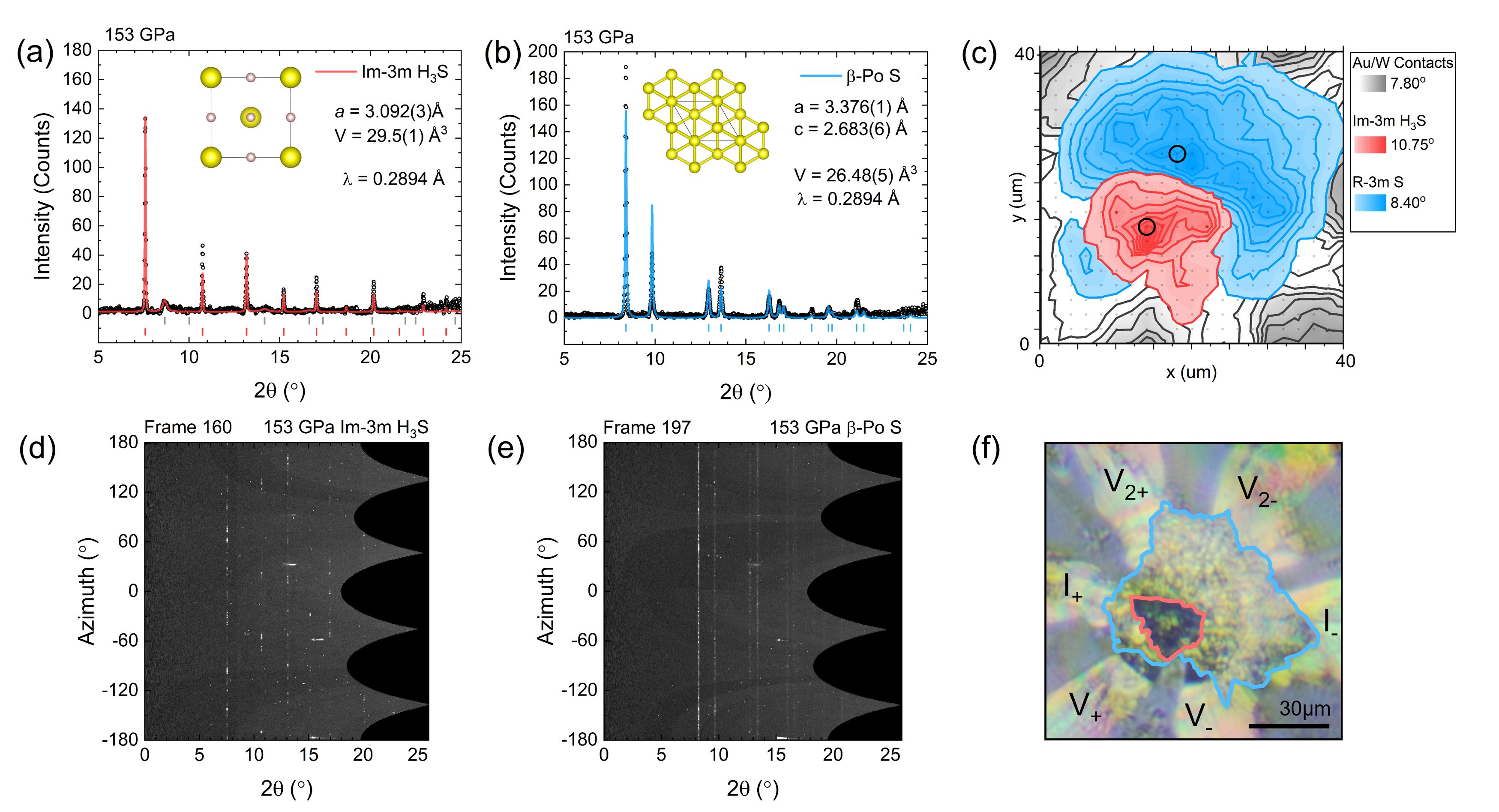}%
\caption{\textbf{Powder x-ray diffraction showing \Imtm\ \HS.} Powder patters taken with a \SI{30}{\second} acquisition time and corrected for diffuse scattering. Panels (a). and  (b). show the integrated patterns at locations where \Imtm\ \HS\ and elemental sulfur (S-V) are respectively present. Vertical ticks indicate expected peak positions for the different phases (gray corresponds to  cubic boron nitride). (c) Spatial mapping of Bragg reflections associated with S-V and \Imtm\ \HS{} phases. Contacts visible in gray, with the S-V and \HS{} regions shown in blue and red respectively. (d) and (e) Unfurled area detector images of powder patterns presented in (a) and (b). Diffuse background correction details are discussed in section SII of the supplementary materials\cite{Supplementary}. (f). Photo of the sample. Labels mark the electrical contacts, while the blue and red lines demarcate the sulfur and \HS{} respectively.} 
\label{fig:XRD}%
\end{figure*}%

Our successful synthesis of \Imtm\ \HS\ is evident from the XRD data collected on our sample at \SI{153}{\GPa} as presented in \autoref{fig:XRD}. After loading sulfur and ammonia borane, we  applied \SI{153}{\GPa} and subsequently laser heated the sample in a small area (cf. section I of the supplemental information for synthesis details\cite{Supplementary, Lord2014}).  In the laser-heated area, we clearly observe an XRD pattern in excellent  agreement with \Imtm\ \HS\ as demonstrated in \autoref{fig:XRD} (a). (see section II of the supplementary information\cite{Supplementary} and references therein\cite{Presher2015_Dioptas, Toby2013_GSASII} for details of the XRD measurements) In particular, we observe no splitting of the (110) Bragg peak at 2$\theta$ = \SI{7.6}{\degree} and hence conclude that a rhombohedral distortion is absent in our sample. This is in agreement with the stability range of the \Imtm\ phase above \SI{140}{\GPa} established from previous XRD measurements \cite{Goncharov2016}. The \Imtm\ phase in our sample consists of larger crystallites compared to the elemental sulfur as evident from the spots in the detector images  \autoref{fig:XRD} (d) and (e). Yet, a preferred orientation appears to be absent as indicated by the good agreement with the Rietveld refinement of the XRD pattern in \autoref{fig:XRD} (a). For the \Imtm\ \HS\ phase, we find a unit cell volume of \SI{29.5(1)}{\angstrom\cubed} in good agreement with previous calculations (\SI{29.2}{\angstrom\cubed})\cite{Duan2014} and with earlier XRD studies at similar pressures (\SI{29.8}{\angstrom\cubed}) where \HS{} is synthesized using elemental precursors \cite{Goncharov2016, Goncharov2017}. Outside the laser-heated area, we find XRD patterns in excellent agreement with elemental sulfur (S-V) in its rhombohedral $\beta$-Po structure (cf. \autoref{fig:XRD} (b)).

Both the XRD mapping and optical image (\autoref{fig:XRD} (c) and (f)) of our sample demonstrate that \Imtm\ \HS\ has formed in an area of $\approx\SI{15}{\um}\times\SI{15}{\um}$. In the optical image, this region is darker than the grainy, metallic elemental sulfur surrounding it and contains a reflective region in its centre. This black region likely marks boron nitride residue from the dissociation of ammonia borane, with weak Bragg reflections associated with boron nitride also visible in \autoref{fig:XRD} (a), whilst the central reflective area constitutes an exposed surface of \Imtm\ \HS\ with metallic reflectivity. Below, we demonstrate that \Imtm\ \HS\ in our sample displays metallic electrical resistance.
The XRD mapping of the characteristic peaks (\autoref{fig:XRD}) confirms that the entire dark region (including the reflective centre) has been transformed to \Imtm\ \HS\, whilst the remainder of the sample is pure elemental sulfur. In addition, we observe the XRD peaks of tungsten and gold from our electrodes at the outer edges of the  area scanned with XRD (gray areas in \autoref{fig:XRD} (c)).

\begin{figure}%
\includegraphics[width=\figurewidth]{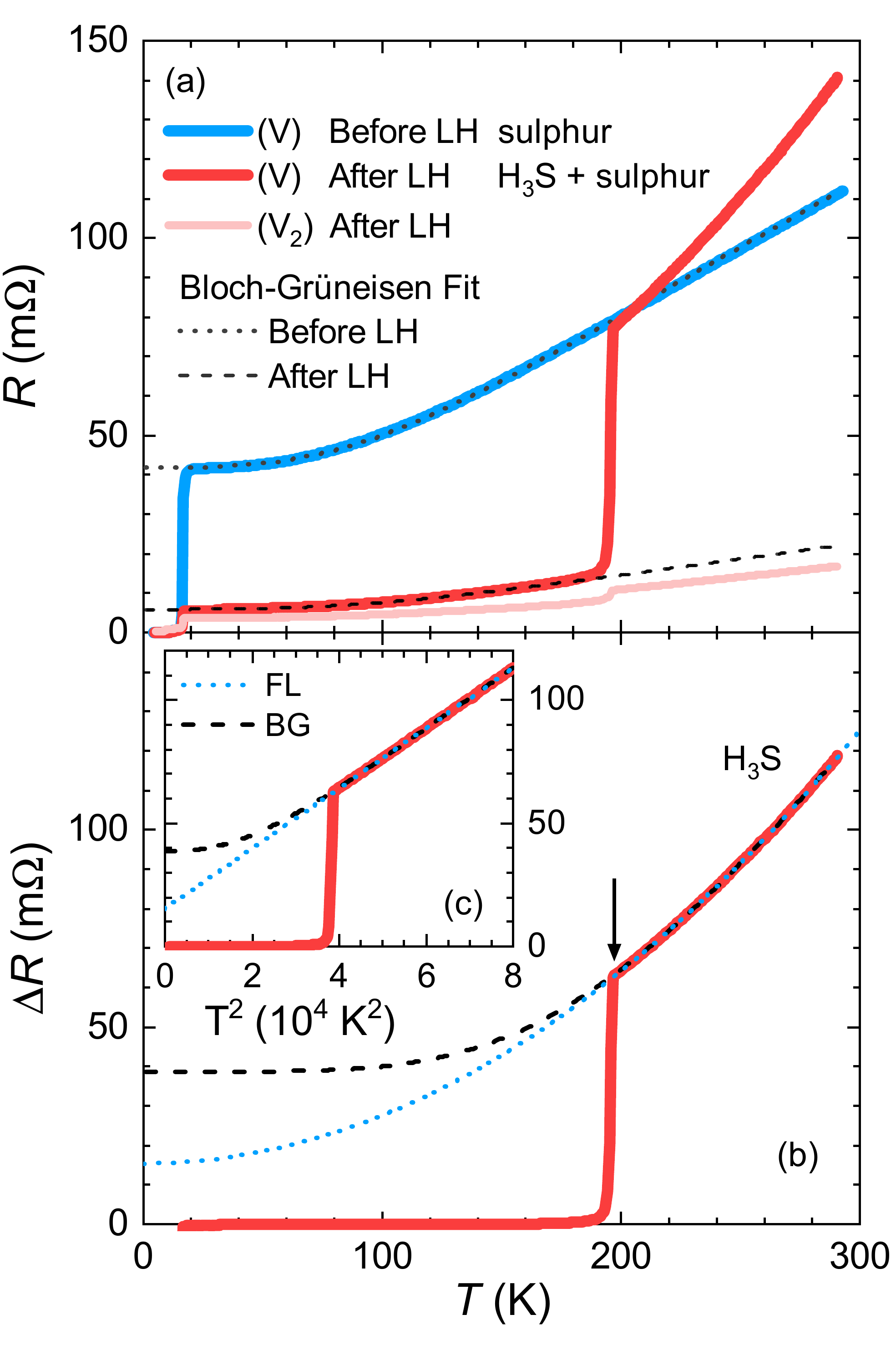}%
\caption{\textbf{High \Tc\ superconducting \HS.} 
(a) Electrical resistance before and after laser heating (LH) measured on two different pairs of contacts (cf. labels in \autoref{fig:XRD} f)). 
The normal resistance of elemental sulfur before laser heating (blue) has been fitted with \autoref{eq:BG} (dotted line) for $\SI{20}{\kelvin}\leq T\leq\SI{290}{\kelvin}$. 
The contribution from sulfur after laser heating has been fitted for $\SI{35}{\kelvin}\leq T\leq \SI{150}{\kelvin}$ with \autoref{eq:BG} (dashed line) using \ThetaD\ and $n$ found for elemental sulfur before laser heating. 
(b) The contribution from \HS\ has been extracted by subtracting the Bloch-Grüneisen (BG) form of the residual sulfur (dashed line in (a)). Vertical arrow marks $\Tc=\SI{197}{\kelvin}$ taken as the intersection between linear fits to the superconducting transition and to the normal state resistance. BG and Fermi-liquid (FL) fits to the normal-state resistance of \HS\ ($T\leq\SI{205}{\kelvin}$) are shown as dashed and dotted lines respectively. 
Inset (c) shows the resistance of \HS\ versus $T^2$ together with both fits.
}
\label{fig:RT}
\end{figure}

Formation of superconducting \Imtm\ \HS\ in our sample is evident from resistivity measurements presented in \autoref{fig:RT}, which show a large  drop in resistance ($\approx\SI{80}{\percent}$) at $\Tc=\SI{197}{\kelvin}$ (see SI III for further details).
Before laser heating, we observe the expected behaviour of elemental sulfur at high pressure: the resistance is metallic and features the superconducting transition of elemental sulfur at \SI{17}{\kelvin} \cite{Struzhkin1997,Gregoryanz2002}.
After laser heating, the resistance shows a major drop at $\Tc=\SI{197}{\kelvin}$ which we associate with the formation of superconducting \HS. Below \Tc, a residual resistance of $\approx\SI{20}{\percent}$ remains which stems from residual elemental sulfur in the measurement path using the contacts $V\pm$. 
With six electrodes, the voltage drop associated with the superconducting transition can be measured in four-point configuration on both sides of the sample.
Using the electrodes (labelled $V\pm$ in \autoref{fig:XRD} f) close to the \Imtm\ \HS\ part of the sample we find an \SI{80}{\percent} drop of the resistance. By contrast, the electrodes further away (labelled $V_2\pm$) yield a drop of less than \SI{25}{\percent}. This shows that superconducting \HS\ is present closer to the electrodes $V\pm$ whilst unreacted sulfur dominates the transport  behaviour sensed between electrodes $V_2\pm$. Hence, we associate the resistive transition with superconductivity of the \Imtm\ phase detected close to the $V\pm$ electrodes.

The resistance contribution originating from \HS\ is extracted in \autoref{fig:RT} (b). Here, the residual resistance from elemental sulfur is subtracted over the full temperature range. For this, we identify the parameters describing the normal state resistance of elemental sulfur from a   Bloch-Grüneisen (BG) fit
\begin{equation}
	R(T) = R_0 + B \left(\frac{T}{\ThetaD}\right)^n\int_0^{\ThetaD/T}{\frac{z^n \dd z}{(e^z-1)(1-e^{-z})}}
\label{eq:BG}
\end{equation}
 between $\SI{20}{\kelvin}\leq T \leq \SI{150}{\kelvin}$ as shown by the dotted line in \autoref{fig:RT} (a). Here, $R_0$ is the residual resistance of the normal state, $B$ quantifies the magnitude of the resistance contribution from electron-phonon scattering, and \ThetaD\ is the Debye temperature. The exponent $n$ assumes different values depending on whether scattering is intra-band ($n=5$) or inter-band ($n=3$) \cite{Ziman1960}. We use parameters specific to elemental sulfur established from fits to $R(T)$ before laser heating as shown by the dotted line in \autoref{fig:RT}: $\ThetaD=\SI{680}{\kelvin}$ and $n=3$. These parameters are in good agreement with recent calculations for elemental sulfur which predicts strong interband scattering \cite{Monni2017}. With \ThetaD\ and $n$ fixed, only $R_0$ and $B$ are fitted to the sulfur contribution in the resistance after laser heating (dashed line in \autoref{fig:RT} (a)). \footnote{We find that the ratio $B/R_0$ is increased by $\approx\SI{40}{\percent}$ for the sulfur contribution after laser heating compared to before laser heating. This suggests that the laser heating has annealed the elemental sulfur and reduced the concentration of dislocations and/or grain boundaries.} We obtain the resistance contribution ($\Delta R$) of \HS\ by subtracting the fit representing the residual sulfur. This corresponds to the assumption of a series-resistor network which is guaranteed below \Tc\ by the superconducting state of \HS\ and likely satisfied above \Tc\ given the proximity of the sensing electrodes to the \HS\ phase in our sample. The resulting curve for the resistance of \HS\ is shown in \autoref{fig:RT} (b).

The normal state behaviour of \Imtm\ \HS\ can be well fitted by the BG form (\autoref{eq:BG}) as shown by the dashed line in \autoref{fig:RT} (b). The limited temperature range f the normal state ($\SI{200}{\kelvin}\leq T \leq\SI{300}{\kelvin}$) does not allow to determine the exponent $n$ for \HS, $n=3$ and $n=5$ yield very similar fits to the data. We fix the exponent to the most common value $n=5$, i.e. assuming that intra-band scattering is dominant in \Imtm\ \HS\ but highlight the resulting uncertainty to other quantities. We find a Debye temperature $\ThetaD=\SI{1260}{\kelvin}$ ($\ThetaD=\SI{1570}{\kelvin}$ for $n=3$) that is considerably larger than for elemental sulfur, reflecting the increased phonon energies stemming from lattice vibrations involving hydrogen. \ThetaD\ is in good agreement with the characteristic phonon frequency calculated by Errea \textit{et al.} \cite{Errea2016}.  We note that the Fermi-liquid (FL) quadratic temperature dependence of the resistance ($R=R_0+A T^2$) discussed in previous work \cite{Mozaffari2019} is consistent with the crossover range of the BG form as shown in the inset \autoref{fig:RT} (c). However, the magnitude of the temperature-dependent resistance is much more consistent with the electron-phonon scattering associated with the BG form than the electron-electron scattering associated with the FL form. Specifically, using the lateral dimensions $\SI{15}{\um} \times \SI{15}{\um}$ and a plausible thickness of \SI{2}{\um} of the \HS\ phase of the sample we estimate the resistivity and the specific magnitude $B^{\prime}=\SI{7e-9}{\ohm\meter}$ of the electron-phonon contribution as well as the specific magnitude of the FL contribution $A^{\prime}=\SI{2.4e-12}{\ohm\meter\per\kelvin\squared}$. Whilst the  uncertainty of $A^{\prime}$ and $B^{\prime}$ is dominated by the uncertainty of the sample thickness $\approx\SI{50}{\percent}$ we can compare to values for other metals. $B^{\prime}$ is comparable in magnitude to values for simple monovalent metals \cite{Ziman1960} 
but $A^{\prime}$ is at least one order of magnitude larger than for transition metals \cite{Rice1968}. 
Hence,  this comparison suggests that the normal-state resistance of \Imtm\ \HS\ is dominated by electron-phonon scattering. 

We estimate the mean free path in our \HS\ sample to be $l\approx\SI{8}{\nano\meter}$. For this, we use the sample dimensions of the \HS\ part to estimate the residual resistivity $\rho_0 \approx \SI{8e-8}{\ohm \meter}$ and the Drude transport equation applied to a single-band free-electron approximation $l=\hbar (3\pi^2)^{1/3}/(n^{2/3}e^2\rho_0)$ with the charge carrier concentration $n\approx \SI{8.5e22}{\per\centi\meter\cubed}$ determined from Hall effect measurements \cite{Mozaffari2019}.

\begin{figure}%
\includegraphics[width=\figurewidth]{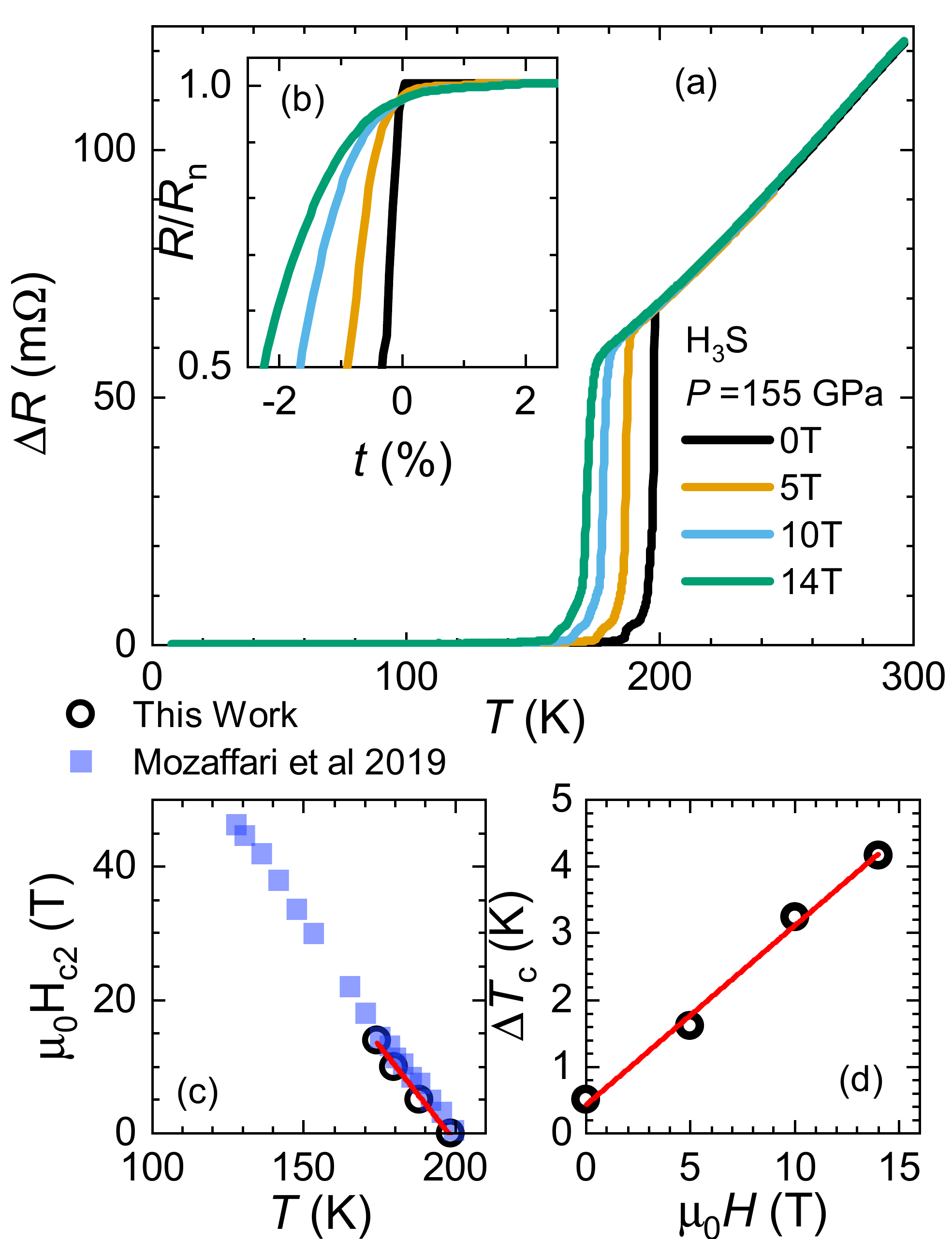}%
\caption{\textbf{Suppression of superconducting transition in magnetic field}. Resistance versus temperature in magnetic fields. Residual resistance has been subtracted following the procedure outlined in \autoref{fig:RT}. Inset (b) shows the resistance normalized to normal state versus reduced temperature $t=(T-\Tc)/\Tc$. (c) shows the upper critical field versus the critical temperature. Here, $\Hcii(T)$ was obtained following $\Tc(H)$ taken as the onset of superconductivity determined as the intersection between the normal state behaviour and a linear fit to the transition. Previous $\Hcii(T)$ data at \SI{155}{\GPa} taken from Mozaffari \textit{et al.} \cite{Mozaffari2019} are included for comparison. (d) shows $\Delta\Tc$, the width of the superconducting transition as a function of magnetic field, given as the difference between \Tc{} onset and the temperature at \SI{50}{\percent} of the normalized drop in resistance. }%
\label{fig:Hc}%
\end{figure}

Further evidence for the superconducting nature of the transition at $\Tc=\SI{197}{\kelvin}$ stems from our measurements in magnetic fields up to \SI{14}{\tesla}. In  \autoref{fig:Hc}(a), we show that the transition is shifted to lower temperature in magnetic field.  The temperature-dependence of the upper critical field, $\Hcii(T)$, associated with superconductivity of \HS\ is in agreement with the behaviour reported earlier by Mozaffari \textit{et al.} (cf.\ \autoref{fig:Hc}(c)) \cite{Mozaffari2019}.  Our samples and those of Mozaffari \textit{et al.} have been synthesized with different methods, from sulfur and ammonia borane or hydrogen respectively. This is very likely to lead to different concentrations of impurities as indicated by the difference of resistance ratios at room temperature and low temperature estimated for the normal-state  $RRR=R(\SI{300}{\kelvin})/R_{\text{normal}}(T\to0)=3$ and $r=1.9$ for our sample and that of Mozaffari \textit{et al.}, respectively.  The fact that we observe good agreement of $\Bcii(T)$ between our samples and  those of Mozaffari \textit{et al.} suggests that the superconducting properties are independent of impurity concentrations, i.e. in the clean limit. By contrast, the samples obtained via dissociation of \HiiS\ by Drozdov \textit{et al.} \cite{Drozdov2015} showed a  much smaller resistance ratio $r\sim1$ and a larger slope at the critical field suggesting an enhancement due to a limited mean free path, i.e. in the dirty limit.

The upper critical field of a superconductor is determined by the orbital and Pauli pair-breaking effects. Near \Tc, the Pauli pair-breaking is negligible for conventional superconductors and hence, we use the slope of the critical field $\dd \Bcii/\dd T = \SI{-0,58(3)}{\tesla\per\kelvin}$ to estimate the coherence length, $\xi_0$, and Fermi velocity, \vF{}. In the clean limit, the slope of the critical field is given by 
\begin{equation}
-\left. \frac{\dd \Bcii}{\dd T}\right|_{\Tc} = \frac{\Phi_0}{2\pi(0.74)^2\xi_0^2\Tc}
\label{eq:xiclean}
\end{equation}
where $\Phi_0=h/2e$ is the flux quantum. From this, we obtain the coherence length $\xi_0 = \SI{2.2}{\nano\meter}$. Comparison of the coherence length with the mean free path ($\xi_0 <l$) justifies applicability of the clean limit relations.
We estimate the Fermi velocity using
\begin{equation}
	\xi_0 =\frac{ \hbar \vF}{\pi \Delta_0} \approx \frac{\hbar \vF}{\pi \gamma \kB \Tc}
\label{eq:vF}
\end{equation}
where the numerical factor $\gamma\approx 2.0(4)$ reflects the expected strong coupling behaviour associated with the high-temperature superconductivity in \HS. Based on these assumptions, we obtain $\vF^{\text{exp}}=\SI{3.8(7)e5}{\meter\per\second}$.

\begin{figure}%
\includegraphics[width=.7\columnwidth]{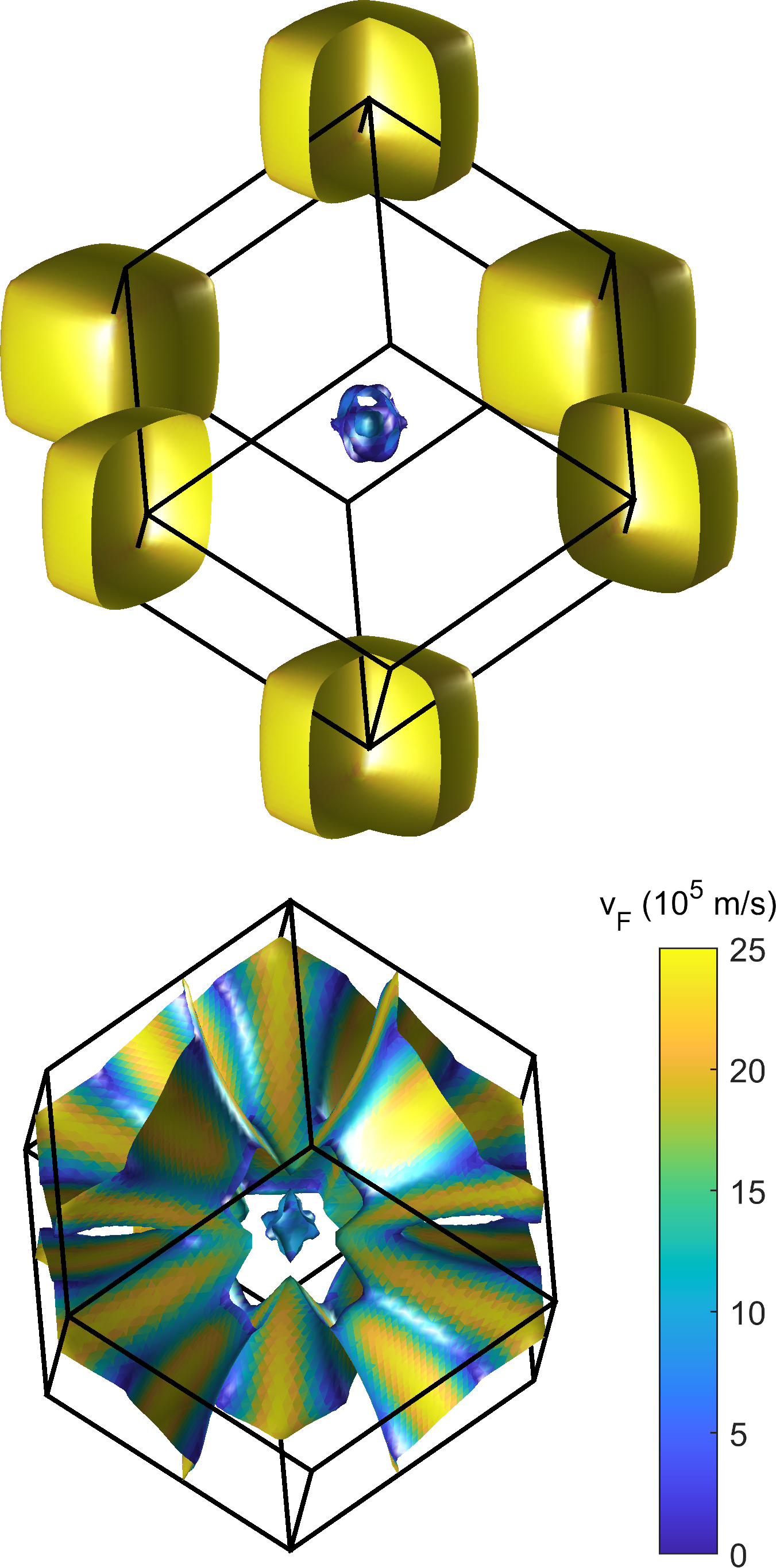}%
\caption{The magnitude of the Fermi velocity is shown as color map  over the Fermi surface of \HS\ as obtained with DFT calculations. Fermi surface sheets 1, 3, and 5 are shown in top panel, sheets 2 and 4 in bottom panel.}%
\label{fig:vF}%
\end{figure}

The obtained Fermi velocity is significantly lower than expected from DFT calculations. We employ Wien2k to calculate the band structure of \HS\ based on the experimental lattice parameters from our XRD results. (Details of the DFT calculation  can be found in section IV of the supplementary information \cite{Supplementary} and references therein\cite{Blaha2019, Perdew1996}.) Our band structure looks very similar to earlier reports \cite{Jarlborg2016}. The Fermi velocity at the Fermi energy is shown as a colormap plot over the Fermi surface in \autoref{fig:vF}. A considerable variation of \vF\ is found over several of the Fermi-surface sheets. Within a simple single-band model we can compare the global average of the Fermi velocity $\langle \vF^{\text{DFT}} \rangle = \SI{1.6e6}{\meter\per\second}$ with our experimental result $\vF^{\text{exp}}\approx\SI{3.8e5}{\meter\per\second}$. The reduced Fermi velocity in experiment compared to DFT confirms strong renormalisation of the Fermi velocity due to the strong electron-phonon coupling in \HS\ \cite{Errea2020}. 

We observe a clear broadening of the superconducting transition in magnetic field. The width of the superconducting transition can be influenced by many factors including inhomogeneities in composition and pressure. In zero field, our sample features a sharp transition with $\Delta \Tc = \Tc^{\text{onset}} - \Tc^{\SI{50}{\percent}} = \SI{0.5}{\kelvin}$ suggesting a good homogeneity of our sample and small pressure gradients. Upon application of a magnetic field, $\Delta\Tc(H)$  increases linearly as demonstrated in \autoref{fig:Hc}(d). 
 A similar linear form with offset has been reported for \YHix\ \cite{Snider2021}. The broadening of the transition is most visible in the inset \autoref{fig:RT}(b). An increase in transition width is in general agreement with expectations for a superconducting state \cite{Hirsch2021}.

\begin{figure}%
\includegraphics[width=\figurewidth]{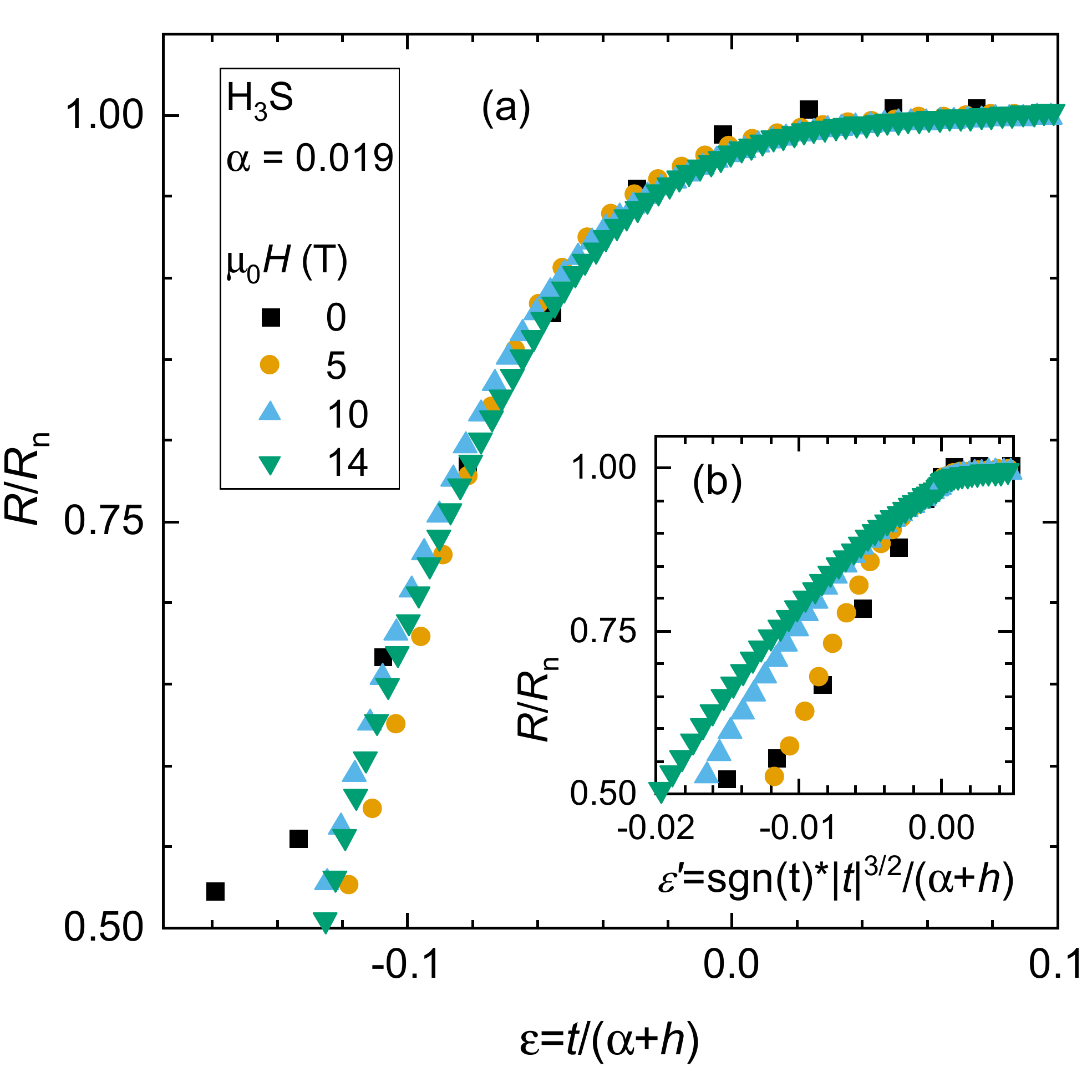}%
\caption{Scaling analysis of the superconducting transition in finite magnetic fields. The resistance normalized to the normal-state resistance determined from Bloch-Grüneisen fits to the data well above \Tc. In (a) $R/R_n$ is shown against $\epsilon=t/(h+\alpha)$ where $t=(T-\Tc)/\Tc$ is the normalized temperature and $h=B/\Bcii(T=0)$ the normalized field. We use $\Bcii=\SI{88}{\tesla}$ as determined by Ref. \cite{Mozaffari2019}. In (b) the normalized resistance is shown against $\epsilon^{\prime} = \sgn(t) |t|^{3/2}/(h+\alpha)$.}%
\label{fig:RB_scaling}%
\end{figure}

In finite magnetic field, flow of the flux lattice can increase the width of the superconducting transition. Near \Tc, the flow of the flux lattice is determined by the activation energy required to overcome pinning leading to the expectation that the resistance scales as  $t^{3/2}/h$ \cite{Tinkham1988}. Here $t=(T-\Tc)/\Tc$ is the normalized temperature and $h=B/\Bcii(T=0)$ the normalized field. Alternatively, the suppression of filamentary superconductivity in finite magnetic fields can cause changes to the superconducting transition -- usually leading to a sharper transition with a lower \Tc.  
By analysing the scaling behaviour of a large portion of the resistive transition, we focus on the contributions arising from the bulk of our \HS\ phase -- the suppression of filamentary superconductivity does not usually lead to a scaling form of the resistance. 
 In our sample of \HS, we find reasonable scaling of a large proportion of the superconducting transition as demonstrated in \autoref{fig:RB_scaling}. Good scaling is found for a functional form  
\begin{equation}
 \frac{R}{R_{\text{normal}}}= f\left(\frac{t^\beta}{h+\alpha}\right) \quad .
	\label{eq:SC}
\end{equation} 
linear in temperature, i.e. with $\beta=1$.
To determine $h$, we use $\Bcii(T=0)=\SI{88}{\tesla}$ estimated by Mozaffari \textit{et al.} \cite{Mozaffari2019} . The constant $\alpha =\num{0.019}$ reflects the finite width of the superconducting transition in zero magnetic field associated with sample inhomogeneities, strain, or other non-thermal causes suggesting that \autoref{eq:SC} implies a   temperature-over-field scaling of the intrinsic superconducting  transition in \HS.
Notably, the scaling of \autoref{eq:SC} leads to much better collapse of the data than a form involving $\beta =3/2$ predicted by Tinkham \textit{et al.} \cite{Tinkham1988} (cf.\ \autoref{fig:RB_scaling}(b)). Whilst a precise determination of $\beta$, i.e. discriminating exponents close to unity will require further measurements over a wider field range the quality of the data collapse in \autoref{fig:RB_scaling} suggests that linear temperature-over-field scaling dominates the fluctuations of the superconducting transition in \HS.
A linear form \autoref{eq:SC} is consistent with data on MgB$_2$ and other hydride superconductors \cite{Cornelius2022}. Yet, we note that the linear transition width and linear temperature-over-field scaling in \HS\ cannot be associated with granular anisotropic superconductivity like in MgB$_2$ \cite{Eisterer2003} as \HS\ is cubic. Hence, the linear temperature-over-field scaling of the resistance in \HS\ suggest that a new model might be required to describe the broadening of the superconducting transition in hydride superconductors.

\section{Conclusion}
\Imtm\ \HS\ can be synthesized from elemental sulfur and ammonia borane. Superconductivity is evident from a sudden drop in resistance which is found to shift to lower temperatures and broaden upon application of magnetic field. Two experimental observations suggest that the superconductivity follows clean-limit behaviour. (i) The critical field curve is found to be identical for different samples at the same pressure synthesized in different ways, i.e. independent of different levels of impurities. (ii) Our estimated mean free path is longer than the coherence length extracted from the slope of the critical field.  A  significant broadening of the superconducting transition in finite field is found and follows a linear $t/(h+\alpha)$ scaling which suggests that a new model is required to describe the fluctuations in proximity to the superconducting transition in hydrides.

\begin{acknowledgments}
The authors thank Antony Carrington for valuable discussions. This work was partially supported by the EPSRC under grants  No. EP/V048759/1 and No. EP/L015544/1, as well as the ERC Horizon 2020 programme under grant 715262-HPSuper. O.L. would like to acknowledge support from the Royal Society in the form of a University Research Fellowship (UF150057). 
We acknowledge DESY (Hamburg, Germany), a member of the Helmholtz Association HGF, for the provision of experimental facilities. Parts of this research were carried out at PETRA III and we would like to thank Hanns-Peter Liermann for assistance in using P02.2. Beamtime was allocated for proposal I-20210376 EC.
\end{acknowledgments}

\section*{Additional information}

Data are available at the University of Bristol data repository \href{https://data.bris.ac.uk/data/}{data.bris} \cite{H3S_RDSF}.

\section*{References}
%


\begin{thebibliography}{43}%
\makeatletter
\providecommand \@ifxundefined [1]{%
 \@ifx{#1\undefined}
}%
\providecommand \@ifnum [1]{%
 \ifnum #1\expandafter \@firstoftwo
 \else \expandafter \@secondoftwo
 \fi
}%
\providecommand \@ifx [1]{%
 \ifx #1\expandafter \@firstoftwo
 \else \expandafter \@secondoftwo
 \fi
}%
\providecommand \natexlab [1]{#1}%
\providecommand \enquote  [1]{``#1''}%
\providecommand \bibnamefont  [1]{#1}%
\providecommand \bibfnamefont [1]{#1}%
\providecommand \citenamefont [1]{#1}%
\providecommand \href@noop [0]{\@secondoftwo}%
\providecommand \href [0]{\begingroup \@sanitize@url \@href}%
\providecommand \@href[1]{\@@startlink{#1}\@@href}%
\providecommand \@@href[1]{\endgroup#1\@@endlink}%
\providecommand \@sanitize@url [0]{\catcode `\\12\catcode `\$12\catcode
  `\&12\catcode `\#12\catcode `\^12\catcode `\_12\catcode `\%12\relax}%
\providecommand \@@startlink[1]{}%
\providecommand \@@endlink[0]{}%
\providecommand \url  [0]{\begingroup\@sanitize@url \@url }%
\providecommand \@url [1]{\endgroup\@href {#1}{\urlprefix }}%
\providecommand \urlprefix  [0]{URL }%
\providecommand \Eprint [0]{\href }%
\providecommand \doibase [0]{https://doi.org/}%
\providecommand \selectlanguage [0]{\@gobble}%
\providecommand \bibinfo  [0]{\@secondoftwo}%
\providecommand \bibfield  [0]{\@secondoftwo}%
\providecommand \translation [1]{[#1]}%
\providecommand \BibitemOpen [0]{}%
\providecommand \bibitemStop [0]{}%
\providecommand \bibitemNoStop [0]{.\EOS\space}%
\providecommand \EOS [0]{\spacefactor3000\relax}%
\providecommand \BibitemShut  [1]{\csname bibitem#1\endcsname}%
\let\auto@bib@innerbib\@empty
\bibitem [{\citenamefont {Drozdov}\ \emph {et~al.}(2015)\citenamefont
  {Drozdov}, \citenamefont {Eremets}, \citenamefont {Troyan}, \citenamefont
  {Ksenofontov},\ and\ \citenamefont {Shylin}}]{Drozdov2015}%
  \BibitemOpen
  \bibfield  {author} {\bibinfo {author} {\bibfnamefont {A.~P.}\ \bibnamefont
  {Drozdov}}, \bibinfo {author} {\bibfnamefont {M.~I.}\ \bibnamefont
  {Eremets}}, \bibinfo {author} {\bibfnamefont {I.~A.}\ \bibnamefont {Troyan}},
  \bibinfo {author} {\bibfnamefont {V.}~\bibnamefont {Ksenofontov}},\ and\
  \bibinfo {author} {\bibfnamefont {S.~I.}\ \bibnamefont {Shylin}},\ }\href
  {https://doi.org/10.1038/nature14964} {\bibfield  {journal} {\bibinfo
  {journal} {Nature}\ }\textbf {\bibinfo {volume} {525}},\ \bibinfo {pages}
  {73} (\bibinfo {year} {2015})}\BibitemShut {NoStop}%
\bibitem [{\citenamefont {Chen}\ \emph {et~al.}(2021)\citenamefont {Chen},
  \citenamefont {Semenok}, \citenamefont {Huang}, \citenamefont {Shu},
  \citenamefont {Li}, \citenamefont {Duan}, \citenamefont {Cui},\ and\
  \citenamefont {Oganov}}]{Chen2021b}%
  \BibitemOpen
  \bibfield  {author} {\bibinfo {author} {\bibfnamefont {W.}~\bibnamefont
  {Chen}}, \bibinfo {author} {\bibfnamefont {D.~V.}\ \bibnamefont {Semenok}},
  \bibinfo {author} {\bibfnamefont {X.}~\bibnamefont {Huang}}, \bibinfo
  {author} {\bibfnamefont {H.}~\bibnamefont {Shu}}, \bibinfo {author}
  {\bibfnamefont {X.}~\bibnamefont {Li}}, \bibinfo {author} {\bibfnamefont
  {D.}~\bibnamefont {Duan}}, \bibinfo {author} {\bibfnamefont {T.}~\bibnamefont
  {Cui}},\ and\ \bibinfo {author} {\bibfnamefont {A.~R.}\ \bibnamefont
  {Oganov}},\ }\href {https://doi.org/10.1103/PhysRevLett.127.117001}
  {\bibfield  {journal} {\bibinfo  {journal} {Phys. Rev. Lett.}\ }\textbf
  {\bibinfo {volume} {127}},\ \bibinfo {pages} {117001} (\bibinfo {year}
  {2021})}\BibitemShut {NoStop}%
\bibitem [{\citenamefont {Snider}\ \emph {et~al.}(2020)\citenamefont {Snider},
  \citenamefont {Dasenbrock-Gammon}, \citenamefont {McBride}, \citenamefont
  {Debessai}, \citenamefont {Vindana}, \citenamefont {Vencatasamy},
  \citenamefont {Lawler}, \citenamefont {Salamat},\ and\ \citenamefont
  {Dias}}]{Snider2020}%
  \BibitemOpen
  \bibfield  {author} {\bibinfo {author} {\bibfnamefont {E.}~\bibnamefont
  {Snider}}, \bibinfo {author} {\bibfnamefont {N.}~\bibnamefont
  {Dasenbrock-Gammon}}, \bibinfo {author} {\bibfnamefont {R.}~\bibnamefont
  {McBride}}, \bibinfo {author} {\bibfnamefont {M.}~\bibnamefont {Debessai}},
  \bibinfo {author} {\bibfnamefont {H.}~\bibnamefont {Vindana}}, \bibinfo
  {author} {\bibfnamefont {K.}~\bibnamefont {Vencatasamy}}, \bibinfo {author}
  {\bibfnamefont {K.~V.}\ \bibnamefont {Lawler}}, \bibinfo {author}
  {\bibfnamefont {A.}~\bibnamefont {Salamat}},\ and\ \bibinfo {author}
  {\bibfnamefont {R.~P.}\ \bibnamefont {Dias}},\ }\href
  {https://doi.org/10.1038/s41586-020-2801-z} {\bibfield  {journal} {\bibinfo
  {journal} {Nature}\ }\textbf {\bibinfo {volume} {586}},\ \bibinfo {pages}
  {373} (\bibinfo {year} {2020})}\BibitemShut {NoStop}%
\bibitem [{\citenamefont {Somayazulu}\ \emph {et~al.}(2019)\citenamefont
  {Somayazulu}, \citenamefont {Ahart}, \citenamefont {Mishra}, \citenamefont
  {Geballe}, \citenamefont {Baldini}, \citenamefont {Meng}, \citenamefont
  {Struzhkin},\ and\ \citenamefont {Hemley}}]{Somayazulu2019}%
  \BibitemOpen
  \bibfield  {author} {\bibinfo {author} {\bibfnamefont {M.}~\bibnamefont
  {Somayazulu}}, \bibinfo {author} {\bibfnamefont {M.}~\bibnamefont {Ahart}},
  \bibinfo {author} {\bibfnamefont {A.~K.}\ \bibnamefont {Mishra}}, \bibinfo
  {author} {\bibfnamefont {Z.~M.}\ \bibnamefont {Geballe}}, \bibinfo {author}
  {\bibfnamefont {M.}~\bibnamefont {Baldini}}, \bibinfo {author} {\bibfnamefont
  {Y.}~\bibnamefont {Meng}}, \bibinfo {author} {\bibfnamefont {V.~V.}\
  \bibnamefont {Struzhkin}},\ and\ \bibinfo {author} {\bibfnamefont {R.~J.}\
  \bibnamefont {Hemley}},\ }\href
  {https://doi.org/10.1103/PhysRevLett.122.027001} {\bibfield  {journal}
  {\bibinfo  {journal} {Phys. Rev. Lett.}\ }\textbf {\bibinfo {volume} {122}},\
  \bibinfo {pages} {027001} (\bibinfo {year} {2019})},\ \Eprint
  {https://arxiv.org/abs/1808.07695v2} {arXiv:1808.07695v2 [cond-mat.mtrl-sci]}
  \BibitemShut {NoStop}%
\bibitem [{\citenamefont {Kong}\ \emph {et~al.}(2021)\citenamefont {Kong},
  \citenamefont {Minkov}, \citenamefont {Kuzovnikov}, \citenamefont {Drozdov},
  \citenamefont {Besedin}, \citenamefont {Mozaffari}, \citenamefont {Balicas},
  \citenamefont {Balakirev}, \citenamefont {Prakapenka}, \citenamefont
  {Chariton}, \citenamefont {Knyazev}, \citenamefont {Greenberg},\ and\
  \citenamefont {Eremets}}]{Kong2021}%
  \BibitemOpen
  \bibfield  {author} {\bibinfo {author} {\bibfnamefont {P.}~\bibnamefont
  {Kong}}, \bibinfo {author} {\bibfnamefont {V.~S.}\ \bibnamefont {Minkov}},
  \bibinfo {author} {\bibfnamefont {M.~A.}\ \bibnamefont {Kuzovnikov}},
  \bibinfo {author} {\bibfnamefont {A.~P.}\ \bibnamefont {Drozdov}}, \bibinfo
  {author} {\bibfnamefont {S.~P.}\ \bibnamefont {Besedin}}, \bibinfo {author}
  {\bibfnamefont {S.}~\bibnamefont {Mozaffari}}, \bibinfo {author}
  {\bibfnamefont {L.}~\bibnamefont {Balicas}}, \bibinfo {author} {\bibfnamefont
  {F.~F.}\ \bibnamefont {Balakirev}}, \bibinfo {author} {\bibfnamefont {V.~B.}\
  \bibnamefont {Prakapenka}}, \bibinfo {author} {\bibfnamefont
  {S.}~\bibnamefont {Chariton}}, \bibinfo {author} {\bibfnamefont {D.~A.}\
  \bibnamefont {Knyazev}}, \bibinfo {author} {\bibfnamefont {E.}~\bibnamefont
  {Greenberg}},\ and\ \bibinfo {author} {\bibfnamefont {M.~I.}\ \bibnamefont
  {Eremets}},\ }\href {https://doi.org/10.1038/s41467-021-25372-2} {\bibfield
  {journal} {\bibinfo  {journal} {Nature Communications}\ }\textbf {\bibinfo
  {volume} {12}},\ \bibinfo {pages} {5075} (\bibinfo {year}
  {2021})}\BibitemShut {NoStop}%
\bibitem [{\citenamefont {Drozdov}\ \emph {et~al.}(2019)\citenamefont
  {Drozdov}, \citenamefont {Kong}, \citenamefont {Minkov}, \citenamefont
  {Besedin}, \citenamefont {Kuzovnikov}, \citenamefont {Mozaffari},
  \citenamefont {Balicas}, \citenamefont {Balakirev}, \citenamefont {Graf},
  \citenamefont {Prakapenka}, \citenamefont {Greenberg}, \citenamefont
  {Knyazev}, \citenamefont {Tkacz},\ and\ \citenamefont
  {Eremets}}]{Drozdov2019}%
  \BibitemOpen
  \bibfield  {author} {\bibinfo {author} {\bibfnamefont {A.~P.}\ \bibnamefont
  {Drozdov}}, \bibinfo {author} {\bibfnamefont {P.~P.}\ \bibnamefont {Kong}},
  \bibinfo {author} {\bibfnamefont {V.~S.}\ \bibnamefont {Minkov}}, \bibinfo
  {author} {\bibfnamefont {S.~P.}\ \bibnamefont {Besedin}}, \bibinfo {author}
  {\bibfnamefont {M.~A.}\ \bibnamefont {Kuzovnikov}}, \bibinfo {author}
  {\bibfnamefont {S.}~\bibnamefont {Mozaffari}}, \bibinfo {author}
  {\bibfnamefont {L.}~\bibnamefont {Balicas}}, \bibinfo {author} {\bibfnamefont
  {F.~F.}\ \bibnamefont {Balakirev}}, \bibinfo {author} {\bibfnamefont {D.~E.}\
  \bibnamefont {Graf}}, \bibinfo {author} {\bibfnamefont {V.~B.}\ \bibnamefont
  {Prakapenka}}, \bibinfo {author} {\bibfnamefont {E.}~\bibnamefont
  {Greenberg}}, \bibinfo {author} {\bibfnamefont {D.~A.}\ \bibnamefont
  {Knyazev}}, \bibinfo {author} {\bibfnamefont {M.}~\bibnamefont {Tkacz}},\
  and\ \bibinfo {author} {\bibfnamefont {M.~I.}\ \bibnamefont {Eremets}},\
  }\href {https://doi.org/10.1038/s41586-019-1201-8} {\bibfield  {journal}
  {\bibinfo  {journal} {Nature}\ }\textbf {\bibinfo {volume} {569}},\ \bibinfo
  {pages} {528} (\bibinfo {year} {2019})}\BibitemShut {NoStop}%
\bibitem [{\citenamefont {Errea}\ \emph {et~al.}(2016)\citenamefont {Errea},
  \citenamefont {Calandra}, \citenamefont {Pickard}, \citenamefont {Nelson},
  \citenamefont {Needs}, \citenamefont {Li}, \citenamefont {Liu}, \citenamefont
  {Zhang}, \citenamefont {Ma},\ and\ \citenamefont {Mauri}}]{Errea2016}%
  \BibitemOpen
  \bibfield  {author} {\bibinfo {author} {\bibfnamefont {I.}~\bibnamefont
  {Errea}}, \bibinfo {author} {\bibfnamefont {M.}~\bibnamefont {Calandra}},
  \bibinfo {author} {\bibfnamefont {C.~J.}\ \bibnamefont {Pickard}}, \bibinfo
  {author} {\bibfnamefont {J.~R.}\ \bibnamefont {Nelson}}, \bibinfo {author}
  {\bibfnamefont {R.~J.}\ \bibnamefont {Needs}}, \bibinfo {author}
  {\bibfnamefont {Y.}~\bibnamefont {Li}}, \bibinfo {author} {\bibfnamefont
  {H.}~\bibnamefont {Liu}}, \bibinfo {author} {\bibfnamefont {Y.}~\bibnamefont
  {Zhang}}, \bibinfo {author} {\bibfnamefont {Y.}~\bibnamefont {Ma}},\ and\
  \bibinfo {author} {\bibfnamefont {F.}~\bibnamefont {Mauri}},\ }\href
  {https://doi.org/10.1038/nature17175} {\bibfield  {journal} {\bibinfo
  {journal} {Nature}\ }\textbf {\bibinfo {volume} {532}},\ \bibinfo {pages}
  {81} (\bibinfo {year} {2016})}\BibitemShut {NoStop}%
\bibitem [{\citenamefont {Errea}\ \emph {et~al.}(2020)\citenamefont {Errea},
  \citenamefont {Belli}, \citenamefont {Monacelli}, \citenamefont {Sanna},
  \citenamefont {Koretsune}, \citenamefont {Tadano}, \citenamefont {Bianco},
  \citenamefont {Calandra}, \citenamefont {Arita}, \citenamefont {Mauri},\ and\
  \citenamefont {Flores-Livas}}]{Errea2020}%
  \BibitemOpen
  \bibfield  {author} {\bibinfo {author} {\bibfnamefont {I.}~\bibnamefont
  {Errea}}, \bibinfo {author} {\bibfnamefont {F.}~\bibnamefont {Belli}},
  \bibinfo {author} {\bibfnamefont {L.}~\bibnamefont {Monacelli}}, \bibinfo
  {author} {\bibfnamefont {A.}~\bibnamefont {Sanna}}, \bibinfo {author}
  {\bibfnamefont {T.}~\bibnamefont {Koretsune}}, \bibinfo {author}
  {\bibfnamefont {T.}~\bibnamefont {Tadano}}, \bibinfo {author} {\bibfnamefont
  {R.}~\bibnamefont {Bianco}}, \bibinfo {author} {\bibfnamefont
  {M.}~\bibnamefont {Calandra}}, \bibinfo {author} {\bibfnamefont
  {R.}~\bibnamefont {Arita}}, \bibinfo {author} {\bibfnamefont
  {F.}~\bibnamefont {Mauri}},\ and\ \bibinfo {author} {\bibfnamefont {J.~A.}\
  \bibnamefont {Flores-Livas}},\ }\href
  {https://doi.org/10.1038/s41586-020-1955-z} {\bibfield  {journal} {\bibinfo
  {journal} {Nature}\ }\textbf {\bibinfo {volume} {578}},\ \bibinfo {pages}
  {66} (\bibinfo {year} {2020})}\BibitemShut {NoStop}%
\bibitem [{\citenamefont {Pickard}\ \emph {et~al.}(2020)\citenamefont
  {Pickard}, \citenamefont {Errea},\ and\ \citenamefont
  {Eremets}}]{Pickard2020}%
  \BibitemOpen
  \bibfield  {author} {\bibinfo {author} {\bibfnamefont {C.~J.}\ \bibnamefont
  {Pickard}}, \bibinfo {author} {\bibfnamefont {I.}~\bibnamefont {Errea}},\
  and\ \bibinfo {author} {\bibfnamefont {M.~I.}\ \bibnamefont {Eremets}},\
  }\href {https://doi.org/10.1146/annurev-conmatphys-031218-013413} {\bibfield
  {journal} {\bibinfo  {journal} {Annu. Rev. Condens. Matter Phys.}\ }\textbf
  {\bibinfo {volume} {11}},\ \bibinfo {pages} {57} (\bibinfo {year}
  {2020})}\BibitemShut {NoStop}%
\bibitem [{\citenamefont {Goncharov}\ \emph {et~al.}(2016)\citenamefont
  {Goncharov}, \citenamefont {Lobanov}, \citenamefont {Kruglov}, \citenamefont
  {Zhao}, \citenamefont {Chen}, \citenamefont {Oganov}, \citenamefont
  {Konôpková},\ and\ \citenamefont {Prakapenka}}]{Goncharov2016}%
  \BibitemOpen
  \bibfield  {author} {\bibinfo {author} {\bibfnamefont {A.~F.}\ \bibnamefont
  {Goncharov}}, \bibinfo {author} {\bibfnamefont {S.~S.}\ \bibnamefont
  {Lobanov}}, \bibinfo {author} {\bibfnamefont {I.}~\bibnamefont {Kruglov}},
  \bibinfo {author} {\bibfnamefont {X.-M.}\ \bibnamefont {Zhao}}, \bibinfo
  {author} {\bibfnamefont {X.-J.}\ \bibnamefont {Chen}}, \bibinfo {author}
  {\bibfnamefont {A.~R.}\ \bibnamefont {Oganov}}, \bibinfo {author}
  {\bibfnamefont {Z.}~\bibnamefont {Konôpková}},\ and\ \bibinfo {author}
  {\bibfnamefont {V.~B.}\ \bibnamefont {Prakapenka}},\ }\href
  {https://link-aps-org.bris.idm.oclc.org/doi/10.1103/PhysRevB.93.174105}
  {\bibfield  {journal} {\bibinfo  {journal} {Phys. Rev. B}\ }\textbf {\bibinfo
  {volume} {93}},\ \bibinfo {pages} {174105} (\bibinfo {year}
  {2016})}\BibitemShut {NoStop}%
\bibitem [{\citenamefont {Li}\ \emph {et~al.}(2016)\citenamefont {Li},
  \citenamefont {Wang}, \citenamefont {Liu}, \citenamefont {Zhang},
  \citenamefont {Hao}, \citenamefont {Pickard}, \citenamefont {Nelson},
  \citenamefont {Needs}, \citenamefont {Li}, \citenamefont {Huang},
  \citenamefont {Errea}, \citenamefont {Calandra}, \citenamefont {Mauri},\ and\
  \citenamefont {Ma}}]{Li2016}%
  \BibitemOpen
  \bibfield  {author} {\bibinfo {author} {\bibfnamefont {Y.}~\bibnamefont
  {Li}}, \bibinfo {author} {\bibfnamefont {L.}~\bibnamefont {Wang}}, \bibinfo
  {author} {\bibfnamefont {H.}~\bibnamefont {Liu}}, \bibinfo {author}
  {\bibfnamefont {Y.}~\bibnamefont {Zhang}}, \bibinfo {author} {\bibfnamefont
  {J.}~\bibnamefont {Hao}}, \bibinfo {author} {\bibfnamefont {C.~J.}\
  \bibnamefont {Pickard}}, \bibinfo {author} {\bibfnamefont {J.~R.}\
  \bibnamefont {Nelson}}, \bibinfo {author} {\bibfnamefont {R.~J.}\
  \bibnamefont {Needs}}, \bibinfo {author} {\bibfnamefont {W.}~\bibnamefont
  {Li}}, \bibinfo {author} {\bibfnamefont {Y.}~\bibnamefont {Huang}}, \bibinfo
  {author} {\bibfnamefont {I.}~\bibnamefont {Errea}}, \bibinfo {author}
  {\bibfnamefont {M.}~\bibnamefont {Calandra}}, \bibinfo {author}
  {\bibfnamefont {F.}~\bibnamefont {Mauri}},\ and\ \bibinfo {author}
  {\bibfnamefont {Y.}~\bibnamefont {Ma}},\ }\href
  {https://link-aps-org.bris.idm.oclc.org/doi/10.1103/PhysRevB.93.020103}
  {\bibfield  {journal} {\bibinfo  {journal} {Phys. Rev. B}\ }\textbf {\bibinfo
  {volume} {93}},\ \bibinfo {pages} {020103} (\bibinfo {year}
  {2016})}\BibitemShut {NoStop}%
\bibitem [{\citenamefont {Huang}\ \emph {et~al.}(2019)\citenamefont {Huang},
  \citenamefont {Wang}, \citenamefont {Duan}, \citenamefont {Sundqvist},
  \citenamefont {Li}, \citenamefont {Huang}, \citenamefont {Yu}, \citenamefont
  {Li}, \citenamefont {Zhou}, \citenamefont {Liu},\ and\ \citenamefont
  {Cui}}]{Huang2019}%
  \BibitemOpen
  \bibfield  {author} {\bibinfo {author} {\bibfnamefont {X.}~\bibnamefont
  {Huang}}, \bibinfo {author} {\bibfnamefont {X.}~\bibnamefont {Wang}},
  \bibinfo {author} {\bibfnamefont {D.}~\bibnamefont {Duan}}, \bibinfo {author}
  {\bibfnamefont {B.}~\bibnamefont {Sundqvist}}, \bibinfo {author}
  {\bibfnamefont {X.}~\bibnamefont {Li}}, \bibinfo {author} {\bibfnamefont
  {Y.}~\bibnamefont {Huang}}, \bibinfo {author} {\bibfnamefont
  {H.}~\bibnamefont {Yu}}, \bibinfo {author} {\bibfnamefont {F.}~\bibnamefont
  {Li}}, \bibinfo {author} {\bibfnamefont {Q.}~\bibnamefont {Zhou}}, \bibinfo
  {author} {\bibfnamefont {B.}~\bibnamefont {Liu}},\ and\ \bibinfo {author}
  {\bibfnamefont {T.}~\bibnamefont {Cui}},\ }\href
  {https://doi.org/10.1093/nsr/nwz061} {\bibfield  {journal} {\bibinfo
  {journal} {Natl Sci Rev}\ }\textbf {\bibinfo {volume} {6}},\ \bibinfo {pages}
  {713} (\bibinfo {year} {2019})}\BibitemShut {NoStop}%
\bibitem [{\citenamefont {Mozaffari}\ \emph {et~al.}(2019)\citenamefont
  {Mozaffari}, \citenamefont {Sun}, \citenamefont {Minkov}, \citenamefont
  {Drozdov}, \citenamefont {Knyazev}, \citenamefont {Betts}, \citenamefont
  {Einaga}, \citenamefont {Shimizu}, \citenamefont {Eremets}, \citenamefont
  {Balicas},\ and\ \citenamefont {Balakirev}}]{Mozaffari2019}%
  \BibitemOpen
  \bibfield  {author} {\bibinfo {author} {\bibfnamefont {S.}~\bibnamefont
  {Mozaffari}}, \bibinfo {author} {\bibfnamefont {D.}~\bibnamefont {Sun}},
  \bibinfo {author} {\bibfnamefont {V.~S.}\ \bibnamefont {Minkov}}, \bibinfo
  {author} {\bibfnamefont {A.~P.}\ \bibnamefont {Drozdov}}, \bibinfo {author}
  {\bibfnamefont {D.}~\bibnamefont {Knyazev}}, \bibinfo {author} {\bibfnamefont
  {J.~B.}\ \bibnamefont {Betts}}, \bibinfo {author} {\bibfnamefont
  {M.}~\bibnamefont {Einaga}}, \bibinfo {author} {\bibfnamefont
  {K.}~\bibnamefont {Shimizu}}, \bibinfo {author} {\bibfnamefont {M.~I.}\
  \bibnamefont {Eremets}}, \bibinfo {author} {\bibfnamefont {L.}~\bibnamefont
  {Balicas}},\ and\ \bibinfo {author} {\bibfnamefont {F.~F.}\ \bibnamefont
  {Balakirev}},\ }\href {https://doi.org/10.1038/s41467-019-10552-y} {\bibfield
   {journal} {\bibinfo  {journal} {Nature Communications}\ }\textbf {\bibinfo
  {volume} {10}},\ \bibinfo {pages} {2522} (\bibinfo {year} {2019})},\ \Eprint
  {https://arxiv.org/abs/1901.11208v1} {1901.11208v1} \BibitemShut {NoStop}%
\bibitem [{\citenamefont {Pace}\ \emph {et~al.}(2020)\citenamefont {Pace},
  \citenamefont {Finnegan}, \citenamefont {Storm}, \citenamefont {Stevenson},
  \citenamefont {McMahon}, \citenamefont {MacLeod}, \citenamefont {Plekhanov},
  \citenamefont {Bonini},\ and\ \citenamefont {Weber}}]{Pace2020}%
  \BibitemOpen
  \bibfield  {author} {\bibinfo {author} {\bibfnamefont {E.~J.}\ \bibnamefont
  {Pace}}, \bibinfo {author} {\bibfnamefont {S.~E.}\ \bibnamefont {Finnegan}},
  \bibinfo {author} {\bibfnamefont {C.~V.}\ \bibnamefont {Storm}}, \bibinfo
  {author} {\bibfnamefont {M.}~\bibnamefont {Stevenson}}, \bibinfo {author}
  {\bibfnamefont {M.~I.}\ \bibnamefont {McMahon}}, \bibinfo {author}
  {\bibfnamefont {S.~G.}\ \bibnamefont {MacLeod}}, \bibinfo {author}
  {\bibfnamefont {E.}~\bibnamefont {Plekhanov}}, \bibinfo {author}
  {\bibfnamefont {N.}~\bibnamefont {Bonini}},\ and\ \bibinfo {author}
  {\bibfnamefont {C.}~\bibnamefont {Weber}},\ }\href
  {https://doi.org/10.1103/PhysRevB.102.094104} {\bibfield  {journal} {\bibinfo
   {journal} {Phys. Rev. B}\ }\textbf {\bibinfo {volume} {102}},\ \bibinfo
  {pages} {094104} (\bibinfo {year} {2020})}\BibitemShut {NoStop}%
\bibitem [{\citenamefont {Guigue}\ \emph {et~al.}(2017)\citenamefont {Guigue},
  \citenamefont {Marizy},\ and\ \citenamefont {Loubeyre}}]{Guigue2017}%
  \BibitemOpen
  \bibfield  {author} {\bibinfo {author} {\bibfnamefont {B.}~\bibnamefont
  {Guigue}}, \bibinfo {author} {\bibfnamefont {A.}~\bibnamefont {Marizy}},\
  and\ \bibinfo {author} {\bibfnamefont {P.}~\bibnamefont {Loubeyre}},\ }\href
  {https://link-aps-org.bris.idm.oclc.org/doi/10.1103/PhysRevB.95.020104}
  {\bibfield  {journal} {\bibinfo  {journal} {Phys. Rev. B}\ }\textbf {\bibinfo
  {volume} {95}},\ \bibinfo {pages} {020104} (\bibinfo {year}
  {2017})}\BibitemShut {NoStop}%
\bibitem [{\citenamefont {Nakao}\ \emph {et~al.}(2019)\citenamefont {Nakao},
  \citenamefont {Einaga}, \citenamefont {Sakata}, \citenamefont {Kitagaki},
  \citenamefont {Shimizu}, \citenamefont {Kawaguchi}, \citenamefont {Hirao},\
  and\ \citenamefont {Ohishi}}]{Nakao2019}%
  \BibitemOpen
  \bibfield  {author} {\bibinfo {author} {\bibfnamefont {H.}~\bibnamefont
  {Nakao}}, \bibinfo {author} {\bibfnamefont {M.}~\bibnamefont {Einaga}},
  \bibinfo {author} {\bibfnamefont {M.}~\bibnamefont {Sakata}}, \bibinfo
  {author} {\bibfnamefont {M.}~\bibnamefont {Kitagaki}}, \bibinfo {author}
  {\bibfnamefont {K.}~\bibnamefont {Shimizu}}, \bibinfo {author} {\bibfnamefont
  {S.}~\bibnamefont {Kawaguchi}}, \bibinfo {author} {\bibfnamefont
  {N.}~\bibnamefont {Hirao}},\ and\ \bibinfo {author} {\bibfnamefont
  {Y.}~\bibnamefont {Ohishi}},\ }\href {https://doi.org/10.7566/JPSJ.88.123701}
  {\bibfield  {journal} {\bibinfo  {journal} {J. Phys. Soc. Jpn.}\ }\textbf
  {\bibinfo {volume} {88}},\ \bibinfo {pages} {123701} (\bibinfo {year}
  {2019})}\BibitemShut {NoStop}%
\bibitem [{\citenamefont {Snider}\ \emph {et~al.}(2021)\citenamefont {Snider},
  \citenamefont {Dasenbrock-Gammon}, \citenamefont {McBride}, \citenamefont
  {Wang}, \citenamefont {Meyers}, \citenamefont {Lawler}, \citenamefont
  {Zurek}, \citenamefont {Salamat},\ and\ \citenamefont {Dias}}]{Snider2021}%
  \BibitemOpen
  \bibfield  {author} {\bibinfo {author} {\bibfnamefont {E.}~\bibnamefont
  {Snider}}, \bibinfo {author} {\bibfnamefont {N.}~\bibnamefont
  {Dasenbrock-Gammon}}, \bibinfo {author} {\bibfnamefont {R.}~\bibnamefont
  {McBride}}, \bibinfo {author} {\bibfnamefont {X.}~\bibnamefont {Wang}},
  \bibinfo {author} {\bibfnamefont {N.}~\bibnamefont {Meyers}}, \bibinfo
  {author} {\bibfnamefont {K.~V.}\ \bibnamefont {Lawler}}, \bibinfo {author}
  {\bibfnamefont {E.}~\bibnamefont {Zurek}}, \bibinfo {author} {\bibfnamefont
  {A.}~\bibnamefont {Salamat}},\ and\ \bibinfo {author} {\bibfnamefont {R.~P.}\
  \bibnamefont {Dias}},\ }\href
  {https://doi.org/10.1103/PhysRevLett.126.117003} {\bibfield  {journal}
  {\bibinfo  {journal} {Phys. Rev. Lett.}\ }\textbf {\bibinfo {volume} {126}},\
  \bibinfo {pages} {117003} (\bibinfo {year} {2021})},\ \Eprint
  {https://arxiv.org/abs/2012.13627} {2012.13627} \BibitemShut {NoStop}%
\bibitem [{\citenamefont {Minkov}\ \emph {et~al.}(2021)\citenamefont {Minkov},
  \citenamefont {Bud'ko}, \citenamefont {Balakirev}, \citenamefont
  {Prakapenka}, \citenamefont {Chariton}, \citenamefont {Husband},
  \citenamefont {Liermann},\ and\ \citenamefont {Eremets}}]{Minkov2021}%
  \BibitemOpen
  \bibfield  {author} {\bibinfo {author} {\bibfnamefont {V.}~\bibnamefont
  {Minkov}}, \bibinfo {author} {\bibfnamefont {S.}~\bibnamefont {Bud'ko}},
  \bibinfo {author} {\bibfnamefont {F.}~\bibnamefont {Balakirev}}, \bibinfo
  {author} {\bibfnamefont {V.}~\bibnamefont {Prakapenka}}, \bibinfo {author}
  {\bibfnamefont {S.}~\bibnamefont {Chariton}}, \bibinfo {author}
  {\bibfnamefont {R.}~\bibnamefont {Husband}}, \bibinfo {author} {\bibfnamefont
  {H.-P.}\ \bibnamefont {Liermann}},\ and\ \bibinfo {author} {\bibfnamefont
  {M.}~\bibnamefont {Eremets}},\ }\bibfield  {journal} {\bibinfo  {journal}
  {Nature Portfolio}\ }\href {https://doi.org/10.21203/rs.3.rs-936317/v1}
  {10.21203/rs.3.rs-936317/v1} (\bibinfo {year} {2021})\BibitemShut {NoStop}%
\bibitem [{\citenamefont {Laniel}\ \emph {et~al.}(2020)\citenamefont {Laniel},
  \citenamefont {Winkler}, \citenamefont {Bykova}, \citenamefont {Fedotenko},
  \citenamefont {Chariton}, \citenamefont {Milman}, \citenamefont {Bykov},
  \citenamefont {Prakapenka}, \citenamefont {Dubrovinsky},\ and\ \citenamefont
  {Dubrovinskaia}}]{Laniel2020}%
  \BibitemOpen
  \bibfield  {author} {\bibinfo {author} {\bibfnamefont {D.}~\bibnamefont
  {Laniel}}, \bibinfo {author} {\bibfnamefont {B.}~\bibnamefont {Winkler}},
  \bibinfo {author} {\bibfnamefont {E.}~\bibnamefont {Bykova}}, \bibinfo
  {author} {\bibfnamefont {T.}~\bibnamefont {Fedotenko}}, \bibinfo {author}
  {\bibfnamefont {S.}~\bibnamefont {Chariton}}, \bibinfo {author}
  {\bibfnamefont {V.}~\bibnamefont {Milman}}, \bibinfo {author} {\bibfnamefont
  {M.}~\bibnamefont {Bykov}}, \bibinfo {author} {\bibfnamefont
  {V.}~\bibnamefont {Prakapenka}}, \bibinfo {author} {\bibfnamefont
  {L.}~\bibnamefont {Dubrovinsky}},\ and\ \bibinfo {author} {\bibfnamefont
  {N.}~\bibnamefont {Dubrovinskaia}},\ }\href
  {https://doi.org/10.1103/PhysRevB.102.134109} {\bibfield  {journal} {\bibinfo
   {journal} {Phys. Rev. B}\ }\textbf {\bibinfo {volume} {102}},\ \bibinfo
  {pages} {134109} (\bibinfo {year} {2020})}\BibitemShut {NoStop}%
\bibitem [{\citenamefont {Buhot}\ \emph {et~al.}(2020)\citenamefont {Buhot},
  \citenamefont {Moulding}, \citenamefont {Muramatsu}, \citenamefont {Osmond},\
  and\ \citenamefont {Friedemann}}]{Buhot2020}%
  \BibitemOpen
  \bibfield  {author} {\bibinfo {author} {\bibfnamefont {J.}~\bibnamefont
  {Buhot}}, \bibinfo {author} {\bibfnamefont {O.}~\bibnamefont {Moulding}},
  \bibinfo {author} {\bibfnamefont {T.}~\bibnamefont {Muramatsu}}, \bibinfo
  {author} {\bibfnamefont {I.}~\bibnamefont {Osmond}},\ and\ \bibinfo {author}
  {\bibfnamefont {S.}~\bibnamefont {Friedemann}},\ }\href
  {https://doi.org/10.1103/PhysRevB.102.104508} {\bibfield  {journal} {\bibinfo
   {journal} {Phys. Rev. B}\ }\textbf {\bibinfo {volume} {102}},\ \bibinfo
  {pages} {104508} (\bibinfo {year} {2020})}\BibitemShut {NoStop}%
\bibitem [{\citenamefont {Einaga}\ \emph {et~al.}(2016)\citenamefont {Einaga},
  \citenamefont {Sakata}, \citenamefont {Ishikawa}, \citenamefont {Shimizu},
  \citenamefont {Eremets}, \citenamefont {Drozdov}, \citenamefont {Troyan},
  \citenamefont {Hirao},\ and\ \citenamefont {Ohishi}}]{Einaga2016}%
  \BibitemOpen
  \bibfield  {author} {\bibinfo {author} {\bibfnamefont {M.}~\bibnamefont
  {Einaga}}, \bibinfo {author} {\bibfnamefont {M.}~\bibnamefont {Sakata}},
  \bibinfo {author} {\bibfnamefont {T.}~\bibnamefont {Ishikawa}}, \bibinfo
  {author} {\bibfnamefont {K.}~\bibnamefont {Shimizu}}, \bibinfo {author}
  {\bibfnamefont {M.~I.}\ \bibnamefont {Eremets}}, \bibinfo {author}
  {\bibfnamefont {A.~P.}\ \bibnamefont {Drozdov}}, \bibinfo {author}
  {\bibfnamefont {I.~A.}\ \bibnamefont {Troyan}}, \bibinfo {author}
  {\bibfnamefont {N.}~\bibnamefont {Hirao}},\ and\ \bibinfo {author}
  {\bibfnamefont {Y.}~\bibnamefont {Ohishi}},\ }\href
  {https://doi.org/10.1038/nphys3760} {\bibfield  {journal} {\bibinfo
  {journal} {Nature Physics}\ }\textbf {\bibinfo {volume} {12}},\ \bibinfo
  {pages} {835} (\bibinfo {year} {2016})}\BibitemShut {NoStop}%
\bibitem [{\citenamefont {Goncharov}\ \emph {et~al.}(2017)\citenamefont
  {Goncharov}, \citenamefont {Lobanov}, \citenamefont {Prakapenka},\ and\
  \citenamefont {Greenberg}}]{Goncharov2017}%
  \BibitemOpen
  \bibfield  {author} {\bibinfo {author} {\bibfnamefont {A.~F.}\ \bibnamefont
  {Goncharov}}, \bibinfo {author} {\bibfnamefont {S.~S.}\ \bibnamefont
  {Lobanov}}, \bibinfo {author} {\bibfnamefont {V.~B.}\ \bibnamefont
  {Prakapenka}},\ and\ \bibinfo {author} {\bibfnamefont {E.}~\bibnamefont
  {Greenberg}},\ }\href {https://doi.org/10.1103/PhysRevB.95.140101} {\bibfield
   {journal} {\bibinfo  {journal} {Phys. Rev. B}\ }\textbf {\bibinfo {volume}
  {95}},\ \bibinfo {pages} {140101} (\bibinfo {year} {2017})}\BibitemShut
  {NoStop}%
\bibitem [{\citenamefont {Villa-Cortés}\ and\ \citenamefont {De~la
  Peña-Seaman}(2022)}]{VillaCortes2022}%
  \BibitemOpen
  \bibfield  {author} {\bibinfo {author} {\bibfnamefont {S.}~\bibnamefont
  {Villa-Cortés}}\ and\ \bibinfo {author} {\bibfnamefont {O.}~\bibnamefont
  {De~la Peña-Seaman}},\ }\href {https://doi.org/10.1016/j.jpcs.2021.110451}
  {\bibfield  {journal} {\bibinfo  {journal} {Journal of Physics and Chemistry
  of Solids}\ }\textbf {\bibinfo {volume} {161}},\ \bibinfo {pages} {110451}
  (\bibinfo {year} {2022})},\ \Eprint {https://arxiv.org/abs/2103.12055}
  {arXiv:2103.12055} \BibitemShut {NoStop}%
\bibitem [{\citenamefont {Troyan}\ \emph {et~al.}(2016)\citenamefont {Troyan},
  \citenamefont {Gavriliuk}, \citenamefont {R{\"{u}}ffer}, \citenamefont
  {Chumakov}, \citenamefont {Mironovich}, \citenamefont {Lyubutin},
  \citenamefont {Perekalin}, \citenamefont {Drozdov},\ and\ \citenamefont
  {Eremets}}]{Troyan2016}%
  \BibitemOpen
  \bibfield  {author} {\bibinfo {author} {\bibfnamefont {I.}~\bibnamefont
  {Troyan}}, \bibinfo {author} {\bibfnamefont {A.}~\bibnamefont {Gavriliuk}},
  \bibinfo {author} {\bibfnamefont {R.}~\bibnamefont {R{\"{u}}ffer}}, \bibinfo
  {author} {\bibfnamefont {A.}~\bibnamefont {Chumakov}}, \bibinfo {author}
  {\bibfnamefont {A.}~\bibnamefont {Mironovich}}, \bibinfo {author}
  {\bibfnamefont {I.}~\bibnamefont {Lyubutin}}, \bibinfo {author}
  {\bibfnamefont {D.}~\bibnamefont {Perekalin}}, \bibinfo {author}
  {\bibfnamefont {A.~P.}\ \bibnamefont {Drozdov}},\ and\ \bibinfo {author}
  {\bibfnamefont {M.~I.}\ \bibnamefont {Eremets}},\ }\href
  {https://doi.org/10.1126/science.aac8176} {\bibfield  {journal} {\bibinfo
  {journal} {Science}\ }\textbf {\bibinfo {volume} {351}},\ \bibinfo {pages}
  {1303} (\bibinfo {year} {2016})}\BibitemShut {NoStop}%
\bibitem [{\citenamefont {Osmond}(2022)}]{Supplementary}%
  \BibitemOpen
  \bibfield  {author} {\bibinfo {author} {\bibfnamefont {I.}~\bibnamefont
  {Osmond}},\ }\href@noop {} {\bibinfo {title} {Supplementary material:
  Clean-limit superconductivity in Im3m H3S synthesized from sulfur and
  hydrogen donor ammonia - borane}} (\bibinfo {year} {2022})\BibitemShut
  {NoStop}%
\bibitem [{\citenamefont {Lord}\ \emph {et~al.}(2014)\citenamefont {Lord},
  \citenamefont {Wann}, \citenamefont {Hunt}, \citenamefont {Walker},
  \citenamefont {Santangeli}, \citenamefont {Walter}, \citenamefont {Dobson},
  \citenamefont {Wood}, \citenamefont {Vočadlo}, \citenamefont {Morard},\ and\
  \citenamefont {Mezouar}}]{Lord2014}%
  \BibitemOpen
  \bibfield  {author} {\bibinfo {author} {\bibfnamefont {O.~T.}\ \bibnamefont
  {Lord}}, \bibinfo {author} {\bibfnamefont {E.~T.~H.}\ \bibnamefont {Wann}},
  \bibinfo {author} {\bibfnamefont {S.~A.}\ \bibnamefont {Hunt}}, \bibinfo
  {author} {\bibfnamefont {A.~M.}\ \bibnamefont {Walker}}, \bibinfo {author}
  {\bibfnamefont {J.}~\bibnamefont {Santangeli}}, \bibinfo {author}
  {\bibfnamefont {M.~J.}\ \bibnamefont {Walter}}, \bibinfo {author}
  {\bibfnamefont {D.~P.}\ \bibnamefont {Dobson}}, \bibinfo {author}
  {\bibfnamefont {I.~G.}\ \bibnamefont {Wood}}, \bibinfo {author}
  {\bibfnamefont {L.}~\bibnamefont {Vočadlo}}, \bibinfo {author}
  {\bibfnamefont {G.}~\bibnamefont {Morard}},\ and\ \bibinfo {author}
  {\bibfnamefont {M.}~\bibnamefont {Mezouar}},\ }\href
  {http://www.sciencedirect.com/science/article/pii/S0031920114001228}
  {\bibfield  {journal} {\bibinfo  {journal} {Physics of the Earth and
  Planetary Interiors}\ }\textbf {\bibinfo {volume} {233}},\ \bibinfo {pages}
  {13} (\bibinfo {year} {2014})}\BibitemShut {NoStop}%
\bibitem [{\citenamefont {Prescher}\ and\ \citenamefont
  {Prakapenka}(2015)}]{Presher2015_Dioptas}%
  \BibitemOpen
  \bibfield  {author} {\bibinfo {author} {\bibfnamefont {C.}~\bibnamefont
  {Prescher}}\ and\ \bibinfo {author} {\bibfnamefont {V.~B.}\ \bibnamefont
  {Prakapenka}},\ }\href {https://doi.org/10.1080/08957959.2015.1059835}
  {\bibfield  {journal} {\bibinfo  {journal} {High Pressure Research}\ }\textbf
  {\bibinfo {volume} {35}},\ \bibinfo {pages} {223} (\bibinfo {year} {2015})},\
  \Eprint {https://arxiv.org/abs/https://doi.org/10.1080/08957959.2015.1059835}
  {https://doi.org/10.1080/08957959.2015.1059835} \BibitemShut {NoStop}%
\bibitem [{\citenamefont {Toby}\ and\ \citenamefont
  {Von~Dreele}(2013)}]{Toby2013_GSASII}%
  \BibitemOpen
  \bibfield  {author} {\bibinfo {author} {\bibfnamefont {B.~H.}\ \bibnamefont
  {Toby}}\ and\ \bibinfo {author} {\bibfnamefont {R.~B.}\ \bibnamefont
  {Von~Dreele}},\ }\href {https://doi.org/10.1107/S0021889813003531} {\bibfield
   {journal} {\bibinfo  {journal} {Journal of Applied Crystallography}\
  }\textbf {\bibinfo {volume} {46}},\ \bibinfo {pages} {544} (\bibinfo {year}
  {2013})}\BibitemShut {NoStop}%
\bibitem [{\citenamefont {Duan}\ \emph {et~al.}(2014)\citenamefont {Duan},
  \citenamefont {Liu}, \citenamefont {Tian}, \citenamefont {Li}, \citenamefont
  {Huang}, \citenamefont {Zhao}, \citenamefont {Yu}, \citenamefont {Liu},
  \citenamefont {Tian},\ and\ \citenamefont {Cui}}]{Duan2014}%
  \BibitemOpen
  \bibfield  {author} {\bibinfo {author} {\bibfnamefont {D.}~\bibnamefont
  {Duan}}, \bibinfo {author} {\bibfnamefont {Y.}~\bibnamefont {Liu}}, \bibinfo
  {author} {\bibfnamefont {F.}~\bibnamefont {Tian}}, \bibinfo {author}
  {\bibfnamefont {D.}~\bibnamefont {Li}}, \bibinfo {author} {\bibfnamefont
  {X.}~\bibnamefont {Huang}}, \bibinfo {author} {\bibfnamefont
  {Z.}~\bibnamefont {Zhao}}, \bibinfo {author} {\bibfnamefont {H.}~\bibnamefont
  {Yu}}, \bibinfo {author} {\bibfnamefont {B.}~\bibnamefont {Liu}}, \bibinfo
  {author} {\bibfnamefont {W.}~\bibnamefont {Tian}},\ and\ \bibinfo {author}
  {\bibfnamefont {T.}~\bibnamefont {Cui}},\ }\href
  {https://doi.org/10.1038/srep06968} {\bibfield  {journal} {\bibinfo
  {journal} {Scientific reports}\ }\textbf {\bibinfo {volume} {4}},\ \bibinfo
  {pages} {6968} (\bibinfo {year} {2014})},\ \bibinfo {note}
  {25382349[pmid]}\BibitemShut {NoStop}%
\bibitem [{\citenamefont {{Struzhkin}}\ \emph {et~al.}(1997)\citenamefont
  {{Struzhkin}}, \citenamefont {{Hemley}}, \citenamefont {{Mao}},\ and\
  \citenamefont {{Timofeev}}}]{Struzhkin1997}%
  \BibitemOpen
  \bibfield  {author} {\bibinfo {author} {\bibfnamefont {V.~V.}\ \bibnamefont
  {{Struzhkin}}}, \bibinfo {author} {\bibfnamefont {R.~J.}\ \bibnamefont
  {{Hemley}}}, \bibinfo {author} {\bibfnamefont {H.-K.}\ \bibnamefont
  {{Mao}}},\ and\ \bibinfo {author} {\bibfnamefont {Y.~A.}\ \bibnamefont
  {{Timofeev}}},\ }\href {https://doi.org/10.1038/37074} {\bibfield  {journal}
  {\bibinfo  {journal} {\nat}\ }\textbf {\bibinfo {volume} {390}},\ \bibinfo
  {pages} {382} (\bibinfo {year} {1997})}\BibitemShut {NoStop}%
\bibitem [{\citenamefont {Gregoryanz}\ \emph {et~al.}(2002)\citenamefont
  {Gregoryanz}, \citenamefont {Struzhkin}, \citenamefont {Hemley},
  \citenamefont {Eremets}, \citenamefont {Mao},\ and\ \citenamefont
  {Timofeev}}]{Gregoryanz2002}%
  \BibitemOpen
  \bibfield  {author} {\bibinfo {author} {\bibfnamefont {E.}~\bibnamefont
  {Gregoryanz}}, \bibinfo {author} {\bibfnamefont {V.~V.}\ \bibnamefont
  {Struzhkin}}, \bibinfo {author} {\bibfnamefont {R.~J.}\ \bibnamefont
  {Hemley}}, \bibinfo {author} {\bibfnamefont {M.~I.}\ \bibnamefont {Eremets}},
  \bibinfo {author} {\bibfnamefont {H.-k.}\ \bibnamefont {Mao}},\ and\ \bibinfo
  {author} {\bibfnamefont {Y.~A.}\ \bibnamefont {Timofeev}},\ }\href
  {https://doi.org/10.1103/PhysRevB.65.064504} {\bibfield  {journal} {\bibinfo
  {journal} {Phys. Rev. B}\ }\textbf {\bibinfo {volume} {65}},\ \bibinfo
  {pages} {064504} (\bibinfo {year} {2002})}\BibitemShut {NoStop}%
\bibitem [{\citenamefont {Ziman}(1960)}]{Ziman1960}%
  \BibitemOpen
  \bibfield  {author} {\bibinfo {author} {\bibfnamefont {J.~M.}\ \bibnamefont
  {Ziman}},\ }\href@noop {} {\emph {\bibinfo {title} {{Electrons and
  phonons}}}}\ (\bibinfo  {publisher} {Clarendon Pr.},\ \bibinfo {address}
  {Oxford},\ \bibinfo {year} {1960})\BibitemShut {NoStop}%
\bibitem [{\citenamefont {Monni}\ \emph {et~al.}(2017)\citenamefont {Monni},
  \citenamefont {Bernardini}, \citenamefont {Sanna}, \citenamefont {Profeta},\
  and\ \citenamefont {Massidda}}]{Monni2017}%
  \BibitemOpen
  \bibfield  {author} {\bibinfo {author} {\bibfnamefont {M.}~\bibnamefont
  {Monni}}, \bibinfo {author} {\bibfnamefont {F.}~\bibnamefont {Bernardini}},
  \bibinfo {author} {\bibfnamefont {A.}~\bibnamefont {Sanna}}, \bibinfo
  {author} {\bibfnamefont {G.}~\bibnamefont {Profeta}},\ and\ \bibinfo {author}
  {\bibfnamefont {S.}~\bibnamefont {Massidda}},\ }\href
  {https://doi.org/10.1103/PhysRevB.95.064516} {\bibfield  {journal} {\bibinfo
  {journal} {Phys. Rev. B}\ }\textbf {\bibinfo {volume} {95}},\ \bibinfo
  {pages} {064516} (\bibinfo {year} {2017})}\BibitemShut {NoStop}%
\bibitem [{Note1()}]{Note1}%
  \BibitemOpen
  \bibinfo {note} {We find that the ratio $B/R_0$ is increased by $\approx \SI
  {40}{\percent }$ for the sulfur contribution after laser heating compared to
  before laser heating. This suggests that the laser heating has annealed the
  elemental sulfur and reduced the concentration of dislocations and/or grain
  boundaries.}\BibitemShut {Stop}%
\bibitem [{\citenamefont {Rice}(1968)}]{Rice1968}%
  \BibitemOpen
  \bibfield  {author} {\bibinfo {author} {\bibfnamefont {M.~J.}\ \bibnamefont
  {Rice}},\ }\href {https://doi.org/10.1103/PhysRevLett.20.1439} {\bibfield
  {journal} {\bibinfo  {journal} {Phys. Rev. Lett.}\ }\textbf {\bibinfo
  {volume} {20}},\ \bibinfo {pages} {1439} (\bibinfo {year}
  {1968})}\BibitemShut {NoStop}%
\bibitem [{\citenamefont {Blaha}\ \emph {et~al.}(2019)\citenamefont {Blaha},
  \citenamefont {Schwarz}, \citenamefont {Madsen}, \citenamefont {Kvasnicka},\
  and\ \citenamefont {Luitz}}]{Blaha2019}%
  \BibitemOpen
  \bibfield  {author} {\bibinfo {author} {\bibfnamefont {P.}~\bibnamefont
  {Blaha}}, \bibinfo {author} {\bibfnamefont {K.}~\bibnamefont {Schwarz}},
  \bibinfo {author} {\bibfnamefont {G.}~\bibnamefont {Madsen}}, \bibinfo
  {author} {\bibfnamefont {D.}~\bibnamefont {Kvasnicka}},\ and\ \bibinfo
  {author} {\bibfnamefont {J.}~\bibnamefont {Luitz}},\ }\href
  {http://www.wien2k.at/} {\emph {\bibinfo {title} {{WIEN2k}}}},\ \bibinfo
  {edition} {19th}\ ed. (\bibinfo {year} {2019})\BibitemShut {NoStop}%
\bibitem [{\citenamefont {Perdew}\ \emph {et~al.}(1996)\citenamefont {Perdew},
  \citenamefont {Burke},\ and\ \citenamefont {Ernzerhof}}]{Perdew1996}%
  \BibitemOpen
  \bibfield  {author} {\bibinfo {author} {\bibfnamefont {J.~P.}\ \bibnamefont
  {Perdew}}, \bibinfo {author} {\bibfnamefont {K.}~\bibnamefont {Burke}},\ and\
  \bibinfo {author} {\bibfnamefont {M.}~\bibnamefont {Ernzerhof}},\ }\href
  {https://doi.org/10.1103/PhysRevLett.77.3865} {\bibfield  {journal} {\bibinfo
   {journal} {Phys. Rev. Lett.}\ }\textbf {\bibinfo {volume} {77}},\ \bibinfo
  {pages} {3865} (\bibinfo {year} {1996})}\BibitemShut {NoStop}%
\bibitem [{\citenamefont {Jarlborg}\ and\ \citenamefont
  {Bianconi}(2016)}]{Jarlborg2016}%
  \BibitemOpen
  \bibfield  {author} {\bibinfo {author} {\bibfnamefont {T.}~\bibnamefont
  {Jarlborg}}\ and\ \bibinfo {author} {\bibfnamefont {A.}~\bibnamefont
  {Bianconi}},\ }\href {https://doi.org/10.1038/srep24816} {\bibfield
  {journal} {\bibinfo  {journal} {Scientific Reports}\ }\textbf {\bibinfo
  {volume} {6}},\ \bibinfo {pages} {24816} (\bibinfo {year}
  {2016})}\BibitemShut {NoStop}%
\bibitem [{\citenamefont {Hirsch}\ and\ \citenamefont
  {Marsiglio}(2021)}]{Hirsch2021}%
  \BibitemOpen
  \bibfield  {author} {\bibinfo {author} {\bibfnamefont {J.~E.}\ \bibnamefont
  {Hirsch}}\ and\ \bibinfo {author} {\bibfnamefont {F.}~\bibnamefont
  {Marsiglio}},\ }\href {https://doi.org/10.1103/PhysRevB.103.134505}
  {\bibfield  {journal} {\bibinfo  {journal} {Phys. Rev. B}\ }\textbf {\bibinfo
  {volume} {103}},\ \bibinfo {pages} {134505} (\bibinfo {year}
  {2021})}\BibitemShut {NoStop}%
\bibitem [{\citenamefont {Tinkham}(1988)}]{Tinkham1988}%
  \BibitemOpen
  \bibfield  {author} {\bibinfo {author} {\bibfnamefont {M.}~\bibnamefont
  {Tinkham}},\ }\href {https://doi.org/10.1103/PhysRevLett.61.1658} {\bibfield
  {journal} {\bibinfo  {journal} {Phys. Rev. Lett.}\ }\textbf {\bibinfo
  {volume} {61}},\ \bibinfo {pages} {1658} (\bibinfo {year}
  {1988})}\BibitemShut {NoStop}%
\bibitem [{\citenamefont {Cornelius}\ \emph {et~al.}(2022)\citenamefont
  {Cornelius}, \citenamefont {Lawler},\ and\ \citenamefont
  {Salamat}}]{Cornelius2022}%
  \BibitemOpen
  \bibfield  {author} {\bibinfo {author} {\bibfnamefont {A.~L.}\ \bibnamefont
  {Cornelius}}, \bibinfo {author} {\bibfnamefont {K.~V.}\ \bibnamefont
  {Lawler}},\ and\ \bibinfo {author} {\bibfnamefont {A.}~\bibnamefont
  {Salamat}},\ }\bibfield  {journal} {\bibinfo  {journal} {arXiv:2202.04254
  [cond-mat.supr-con]}\ }\href {https://doi.org/10.48550/arXiv.2202.04254}
  {10.48550/arXiv.2202.04254} (\bibinfo {year} {2022}),\ \Eprint
  {https://arxiv.org/abs/2202.04254} {arXiv:2202.04254 [cond-mat.supr-con]}
  \BibitemShut {NoStop}%
\bibitem [{\citenamefont {Eisterer}\ \emph {et~al.}(2003)\citenamefont
  {Eisterer}, \citenamefont {Zehetmayer},\ and\ \citenamefont
  {Weber}}]{Eisterer2003}%
  \BibitemOpen
  \bibfield  {author} {\bibinfo {author} {\bibfnamefont {M.}~\bibnamefont
  {Eisterer}}, \bibinfo {author} {\bibfnamefont {M.}~\bibnamefont
  {Zehetmayer}},\ and\ \bibinfo {author} {\bibfnamefont {H.~W.}\ \bibnamefont
  {Weber}},\ }\href {https://doi.org/10.1103/PhysRevLett.90.247002} {\bibfield
  {journal} {\bibinfo  {journal} {Phys. Rev. Lett.}\ }\textbf {\bibinfo
  {volume} {90}},\ \bibinfo {pages} {247002} (\bibinfo {year}
  {2003})}\BibitemShut {NoStop}%
\bibitem [{\citenamefont {Friedemann}(2022)}]{H3S_RDSF}%
  \BibitemOpen
  \bibfield  {author} {\bibinfo {author} {\bibfnamefont {S.}~\bibnamefont
  {Friedemann}},\ }\href@noop {} {\bibinfo {title} {{Data for publication
  “Clean-limit superconductivity in Im3m H3S synthesized from sulfur and
  hydrogen donor ammonia borane"}}},\ \bibinfo {howpublished}
  {\doi{10.5523/bris.31o8e84oir4ug21mx9vqymsjyz}} (\bibinfo {year}
  {2022})\BibitemShut {NoStop}%
\end{thebibliography}%

\end{document}


\title{Supplemental Material: Clean-limit superconductivity in \Imtm\ \HS\ synthesised from sulphur and hydrogen donor ammonia borane}

\author{Israel Osmond}
\affiliation{H.H. Wills Physics Laboratory, University of Bristol, Tyndall Avenue, Bristol, BS8 1TL, UK}

\author{Owen Moulding}
\affiliation{H.H. Wills Physics Laboratory, University of Bristol, Tyndall Avenue, Bristol, BS8 1TL, UK}

\author{Sam Cross}
\affiliation{H.H. Wills Physics Laboratory, University of Bristol, Tyndall Avenue, Bristol, BS8 1TL, UK}

\author{Takaki Muramatsu}
\affiliation{H.H. Wills Physics Laboratory, University of Bristol, Tyndall Avenue, Bristol, BS8 1TL, UK}

\author{Annabelle Brooks}
\affiliation{H.H. Wills Physics Laboratory, University of Bristol, Tyndall Avenue, Bristol, BS8 1TL, UK}

\author{Oliver Lord}
\affiliation{School of Earth Sciences, University of Bristol, Wills Memorial Building, Queen’s Road,
Bristol BS8 1RJ, United Kingdom}

\author{Timofey Fedotenko}
\affiliation{ Photon Science, DESY, Notkestrasse 85, 22607 Hamburg, Germany}

\author{Jonathan Buhot}
\affiliation{H.H. Wills Physics Laboratory, University of Bristol, Tyndall Avenue, Bristol, BS8 1TL, UK}

\author{Sven Friedemann}
\affiliation{H.H. Wills Physics Laboratory, University of Bristol, Tyndall Avenue, Bristol, BS8 1TL, UK}
\email{Sven.Friedemann@bristol.ac.uk}
	
\date{\today}
\maketitle

\section{Sample Synthesis}
High-pressure measurements were carried out using home-designed diamond anvil cells equipped with diamonds with \SI{50}{\um} culet. The gasket was made from T301 stainless steel, preindented to a thickness below \SI{40}{\um}. The centre of the steel was then drilled out and replaced with a mixture of cubic-boron nitride and stycast epoxy 2850 FT to insulate the electrical leads from the gasket.

Sulphur (\SI{99.5}{\percent}, Alfa Aesar) was flattened to approximately \SI{2}{\um} thick and then cut by hand to form a sample of $\approx \SI{20}{\um}\times\SI{20}{\um}\times\SI{2}{\um}$. 
Elemental sulphur was loaded together with ammonia borane (NH$_3$BH$_3$, 97\%, Sigma Aldrich), which acted as a quasi-hydrostatic pressure medium and hydrogen donor. 
After pressurising to \SI{153}{\GPa}, the  elemental sulphur sample and the ammonia borane  was heated by $\sim$20 pulses from a Yb-YAG fibre laser with a wavelength of \SI{1070}{\nano\meter}, focused to a diameter of $\sim\SI{20}{\micro\meter}$ close to the centre of the culet. Each pulse had a duration of \SI{0.15}{\second}, with power ranging from \SIrange{32}{36}{\watt}. Temperature was measured during each pulse using standard spectroradiometric techniques, with visible changes to the sample happening at a synthesis temperature $\approx\SI{2000}{\kelvin}$. The laser heating and temperature measurement system employed, and its calibration, is detailed in \cite{Lord2014} with the following differences: beam-shaping optics were not used in this study and all 256 rows of the CCD used to collect the incandescent light were binned to improve the signal to noise ratio. This was necessary due to the very short acquisition time needed to prevent the destruction of the diamonds by thermal stresses, with the inevitable consequence being a loss of spatial information.

We note that we had $\approx 10$ failed attempts to synthesise superconducting \HS\ from dissociation of cryogenically loaded H$_2$S following various $p-T$ paths including compression at liquid nitrogen temperature. 

\section{X-ray diffraction}
Synchrotron XRD data were collected at the P02.2 beamline at the Petra synchrotron in DESY. Angle dispersive X-ray diffraction patterns were collected for incident wavelength of $\lambda = \SI{0.2894}{\angstrom}$ and recorded on a PerkinElmer XRD 1621 plate detector. All powder patterns were taken at room temperature with \SI{30}{\second} acquisition times. In order to correct for the diffuse scattering, each area detector image was subject to a pixel-by-pixel background subtraction process. Spatial mapping was performed by rastering over a $\SI{40}{\micro\meter}\times\SI{40}{\micro\meter}$ grid centred on the sample. The minimum recorded intensity for a given pixel across all positions was collated to form a background image representing  the diffuse background. This collated image of the background was carefully checked not to contain any ring-like structures indicating residual XRD from the sample. Area detector images after subtraction of the background were  integrated in DIOPTAS \cite{Presher2015_Dioptas} to yield intensity vs $2\theta$ patterns. Structural refinements to these patterns were  performed in GSAS-II \cite{Toby2013_GSASII}. Presented refinements are all of Reitveld type, with no additional texture included.

\section{Electrical Transport}
Transport measurements were performed using six bilayer electrodes deposited onto the diamond anvil. These electrodes were formed from sputtering \SI{150}{\nano\meter} of tungsten and evaporating \SI{50}{\nano\meter} of gold through a shadow mask. 

Electrical resistance measurements were done in four-point configuration, using  an  ac-resistance bridge  SIM921,  Stanford  Research  Systems  with  \SI{100}{\micro\ampere} excitation current. Resistance measurements under applied magnetic field were performed with a SR830 (Stanford Research Systems) lock-in amplifier.

\section{Density-functional Calculations}
We use Wien2k for our density-functional theory (DFT) calculations \cite{Blaha2019} with the PBE exchange functional \cite{Perdew1996}, an $\text{RKmax}=6$, and an energy range $-6$ to \SI{5}{\rydberg} which corresponds to a valence band treatment of the 1s orbitals for hydrogen and 3s and 3p states for sulphur. Relativistic effects and spin-orbit coupling are ignored in this computation owing to the light masses of hydrogen and sulphur. The calculations are converged for 286 points in the irreducible wedge. The band structure was subsequently evaluated for 2444 points and used to calculate the Fermi velocity from the gradients at the Fermi energy.

\bibliography{bibfile}